# Article

open

# Light's Impact on Fertility: Unveiling the Maybe Connection Between Nighttime Illumination and Global Societal Changes

Wenqing Fang[1]†, Chaopu Yang[1,2]†, Taiyang Chen[1], Youming Zhang[1], Shanxiao Huang[1], Kaiqi Fang[1], Tuanqing Fang[1], Bo Yang[1], Xudong Wang[1]†, Chaoliao[1]

All China is talking about its lowest crude birth rate in 2023 (6.4‰) in the world. It may be the low fertility rate in advanced countries that had led to the transfer of productivity and technology to China, which is triggering a "New Camp Rivalry". Whether all this is related to the wrong use of various artificial night light rays passing through our eyes is worth human scientific reflectio.Non-visual effects of light are the effects of the blue component of light, which effectively inhibits the melatonin secretion in the pineal gland[1–4]. As melatonin acts through high-affinity receptors located centrally and in all organs[5], blue light affects, in principle, the hormone secretion throughout the body, including the secretion of sex hormones and cortisol[5–6]. These effects have been widely used in animal reproduction[7–9]. However, blue light hazard usually focuses on the impact on eyes, rhythm, and sleep, or at most, on obesity, diabetes, and breast cancer related with endocrine disorders[10–11]. Given that the night light environment has significantly changed over the past century, an effect of light on fertility and behavior could be detected in multiple cases. We showed that the ovulation phase of hens could be shifted by the light as low as 0.2 Lx, which corresponds to the illumination intensity at full moon,and **the effect of 0.2Lx on ovulation is normally far greater than all other factors combined**.Two rounds of experiments and many other facts listed support that human intrinsic fertility may have not declined significantly, and food, chemical pollution and police making are all not the main factors affecting fertility. The longer lighting time at night has brought modern humans into a radically different endocrine state compared with that of humans who lived 100 years ago, which has a huge impact on human reproduction, policy making, values and even the progress of civilization. With the development of the digital era, our eyes will be exposed to even longer light at night. How to restore human beings to a normal hormonal state by combating the misuse of light may be the urgent issue of concern to every country, not lunching a new camp rivalry. The light of the digital world (or Metaverse, or AI, or any high technology）should not continue to flood. Modern humans should not be in the wrong hormonal state to pursue freedom, to earn more money and make decisions.

**Light exposure at night has changed significantly in the last century.** In 2019, the International Commission on Illumination (CIE) published the CIE Position Statement on Non-Visual Effects of Light and recommended proper light at the proper time[1]. This raises the questions which lighting mode is proper and what is the worst effect because humans haven't used light properly for a hundred years? In 1879, Edison invented the incandescent light bulb, and since then, electric light at night has been regarded as a sign of modern civilization. However, it has to be questioned whether this evaluation is still correct. Before Edison's invention, most humans used firelight as night light. The unit of illumination, Lx, refers to the illumination intensity of candlelight in a distance of 1 m. About 60 years ago, in rural China, a family shared only one oil lamp or candle at night. Children were only exposed to an amount of illumination of 1 Lx during reading or writing at a distance of 1 m from the oil lamp, while spinning yarn at a distance of 5 m resulted in an exposure to only 0.04 Lx. Nowadays, classroom illumination is 400 Lx, which is $10^4$ times higher than that of spinning yarn 60 years ago. Considering that the content of blue light in white lighting is 12 times higher than that of candlelight (Extended Data (ED) Table 1), the non-visual effects of light in artificial light sources have increased dramatically by 120,000 times over the past century (without considering the change of pupil size with light). Furthermore, the photoperiod has increased significantly from 12 h to 17 h. The effects of the increase in non-visual effects and photoperiod of light on human fertility, consciousness, and behavior have not been explored to date.

**Influence of light on the endocrine function in vertebrates**. All four endocrine axes, including the pineal, reproductive, adrenal, and thyroid axes, could be affected by artificial light at night[5]. Light is widely used as animal aphrodisiac and ripenin in animal farms[7–9], and night light should be considered a new endocrine disruptor[12,32]. However, no studies on the impact of light on fertility, behavior, and values have been carried out due to the complex factors in humans. Therefore, it has to be investigated if the significant changes in the photoperiod and intensity of light that occurred over the past century led to an abnormal state of (sexual) excitement of modern humans for decades. Fig. 1 shows the deductive reasoning and structure of this paper.

## Huge effect of light on animal reproduction (E2 in Fig.1)

**Reversal of the egg-laying phase in nine days by light with an intensity of 0.2 Lx** (Fig. 2). The ground illumination during full moon is 0.2 Lx. A hen usually lays an egg between 10:00 am and 13:00 pm each day. To stimulate hens to lay eggs at night, the lighting must be adjusted to allow hens rest in the dark during the day and be exposed to light at night. We designed an egg-counting method to study the effects of illumination, brightness, dark, light spectrum, "working in three shift", eating start time, and the sound in the dark on the laying phase without

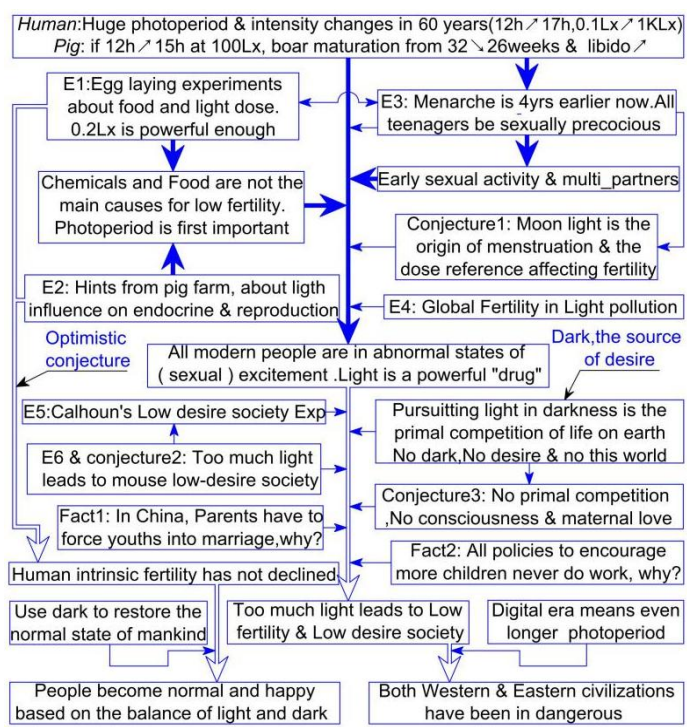

**Fig. 1 | Structure of this paper.** Large solid arrows point to the main conclusions or discovers we try to prove; Large hollow arrows point to the main conjectures worthy of the highest human attention; "→" indicates the main evidences or correlations we could find. E1- E6 represent six experiments, of which, E1 done by us was used to prove that the photoperiod of 0.2 Lx weak light be far powerful for egg laying phase than food. E2- E5 could be seen as the super large scale and long term experiments for this paper only after food and chemical factors were ruled out.

[1]Institute of material Science, Nanchang University, 235# NanJinDong Road, Nanchang, JIangxi 330046,P.R.China. [2] School of Chemistry and Materials, Shangluo College, Shangluo, Shaanxi，P.R.China．
†Present addresses: fwq@ncu.edu.cn，cpyang2022@sinano.ac.cn

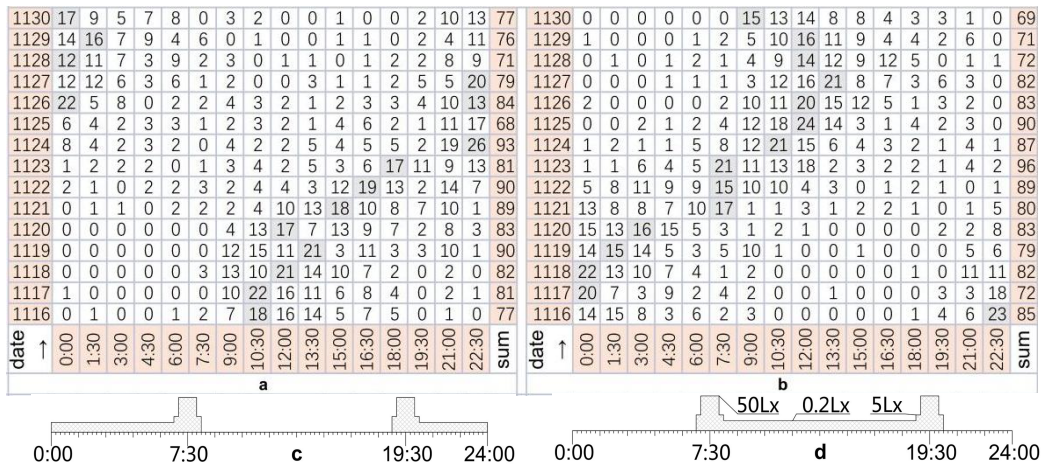

**Fig. 2 | Light at a low intensity of 0.2 Lx (≈ ground illumination at full moon) can completely shift the egg-laying phase about 9 days.** The vertical axis is the test date (e.g., 1116 stands for 2019-11-16), and the horizontal axis indicates the hours of a day (24 h are divided into 16 time periods of 1.5 h each). Infrared cameras were turned on every 1.5 h for 10 s to record the number of eggs laid during the past 1.5 h. The maximum egg production each day is highlighted in grey. The 208 hens were divided evenly into flocks A and B, which were raised in hen houses **a** and **b**. T (25 ± 0.5 °C) and food supply (total 90 min, starting at 6:45 am & 18:45 pm) were the same in houses **a** and **b**. During feeding, the light conditions in houses **a** and **b** were also identical and consisted of three segments (15 min (5 Lx) + 60 min (50 Lx) + 15 min (5 Lx)). In contrast, **c** and **d** show the inversed light conditions that were realized in houses **a** and **b** at other times. The pre-test conditions were realized by using 10 Lx light to approach a natural light rhythm in flock A, while opposite conditions were applied in flock B. After hen house A was illuminated at night, it took about 9 days (1116–1124) for the egg production peak to shift from around 10:30 to 22:30, while the opposite effect was observed in house B .(**sum**, total number of eggs per day).

disturbing the hens (see **Methods** and ED Fig. 1- 8 for details). Several novel conclusions are listed in the Extended Data, and the main conclusions are summarized in the following: 1) The light intensity during full moon (0.2 Lx) is strong enough to affect ovulation. 2) A light intensity of 10 Lx is close to its saturation value for all diurnal animals, and no essential difference between the effects of 0.2 Lx (moonlight) and $10^5$ Lx (sunlight) on ovulation were found. The effect of light intensity on ovulation is less important than that of the photoperiod or the number of hours of light per day. 3) Both the "dark test" and "dietary start time" test showed that food had no effect on ovulation as long as nutrition was guaranteed, which also indicates that the effect of chemical pollution on fertility should be less important than expected.Otherwise, the egg peak in **Fig.2** should disappear first or a peak corresponding to other factors should appear

**Exposure of hens to "three shifts working" light (ED Fig. 2 and Method (M) 21).** Exposure of hens to three lighting intervals resulted in disordered egg-laying phases, which suggests that women of reproductive age maintain stable photoperiod in order to improve their fertility. We concluded that the order of effect of light factors on ovulation was photoperiod, spectrum, intensity, stability and focusing, while dietary factors and chemical pollution do usually not affect human reproduction. These factors are often not distinguished correctly.

**Probable control of the menstruation cycle by moonlight changes during one month.** According to the 2017 Nobel Prize in Physiology or Medicine, it can be inferred that menstruation is a genetically determined human physiological rhythm whose phase regulation should also be related to light. We support the hypothesis that the rhythm of the menstruation cycle originates from changes in the moonlight over a month rather than the change of the tide height in a month (**M27, M2**). In fact, no indicator in humans is related to the moon's gravitational changes. As light at the intensity of 0.2 Lx can control the egg-laying phase of hens, we may assume that that the full-moon light intensity of 0.2 Lx affects ovulation in humans. Compared with this light intensity, today's powerful artificial lighting at nighttime, such as the common 300 Lx desktop illumination, is up to 1,500 times more powerful.

**An empirical formula explains the mismatch between menstruation and moonlight in modern women.** The hen illumination test showed that it may take the time nT to reverse the egg-laying phase (T, period of the biological clock; n ≥ 3, n is determined by the light intensity in ED Fig. 4). From T = 1 d (day) and n = 9 for hens at 0.2 Lx (n= 6 at 3 Lx , n=3 at 10 Lx), it follows T= 30 d, n= 9, and nT= 270 d for a woman to reverse her menstruation phase at moonlight (0.2 Lx). The mismatch (see **M27** for details) can be explained by the humans' use of fire: firelight is both stronger and more irregular than moonlight, and it is too complex to synchronize the menstruation with moonlight for a long period of 270 days. Therefore, considering the influence of light on human hormone secretion, firelight might have induced the first huge leap in their reproductive system (Fig. 3; the 2nd and 3rd leaps may be caused by the use of electric lamps and smartphones respectively). It is necessary to assess if night lighting after the Industrial Revolution has had a significant impact on human menstruation and reproduction.We will discuss specifically whether menstrual synchrony exists in a separate paper(**M27**).

**Light may cause an earlier menarche (E3 in Fig.1)**. The onset of menarche is preceded by a complex cascade of hormonal changes during puberty. The early onset of pubertal development (Fig. 3) is an important medical and social problem, as it may be associated with health problems in later life, such as breast cancer, type 2 diabetes, fertility impairment, cardiovascular diseases, obesity, psychological disorders, high risk of deviant behavior, and early sexual debut[13]. Various factors that decrease the age at menarche have been studied, including genetic and non-genetic factors, as well as prenatal and postnatal factors[13–15], but no factors related to lighting have been studied yet. Nutrition and environmental hazards are generally considered to be the most important factors. However, we proved that these hazards are not the most relevant factors and hypothesized that light affects endogenous hormones, which cause girls to overeat. Environmental hormones are exogenous factors, and their effects are not as straightforward as those of light (**M2**), which may be the main factor that causes central precocious puberty. Compared to people who lived 100 years ago, people nowadays are generally precocious, which may indicate psychological disorder or deviant behavior.

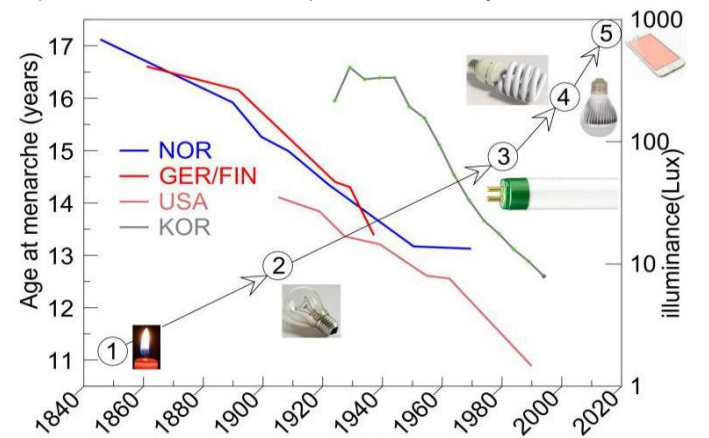

**Fig. 3 | Possible relationship between the significant decline in the age at menarche and night lighting innovations in the past 150 years**[15–17] **("God's lighting experiment with human subjects").** As humans sense light intensity logarithmically according to Optics, the logarithm of the illumination in Lux is plotted on the vertical axis on the right to display the desktop illumination by five types of lamps and the eras of their highest popularity. According to measurements, desktop illumination by candles is about 3 Lx, by incandescent light bulbs about 10 Lx, by early fluorescent lamps about 90 Lx, by rare earth fluorescent lamps about 400 Lx, and by LED lamps (or by smartphones equivalently) about 750–1000 Lx. Since the invention of the incandescent light bulb in 1879, the age at menarche in advanced countries declined evidently. The age at menarche in South Korea only began to decline in the 1950s due to the late adoption of electric lighting. This difference between South Korea and technologically further advanced countries confirms that the earlier menarche may be related to the exposure to more light at night.

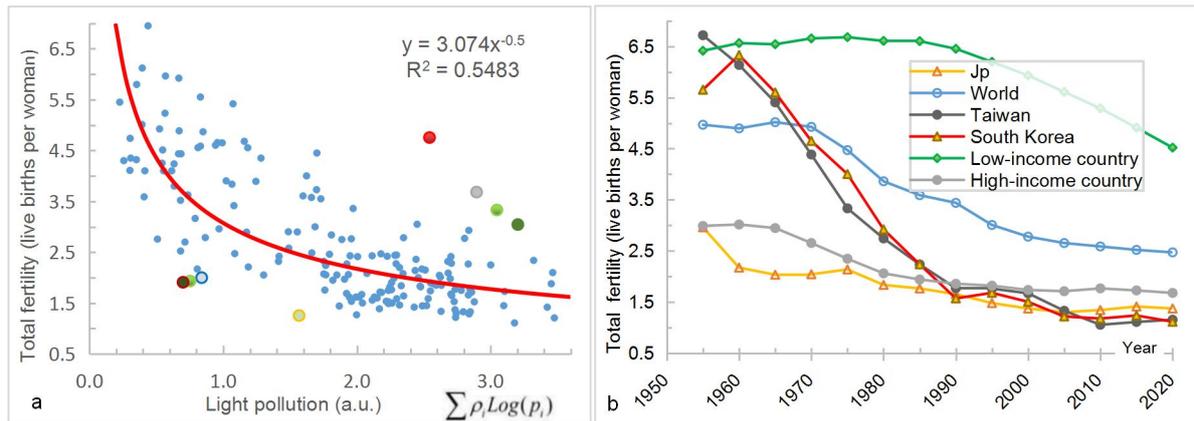

**Fig. 4 | Summary of the effects of light on the fertility ("God's Experiment") in 184 countries over 70 years (each dot stands for one country). a |** Higher light pollution indicates roughly longer photoperiods and stronger light, which may have induced lower fertility in countries with abundant nighttime light. The modified light pollution data[27] (**M26**) was related with the world's fertility data in 2019 published by the UN[26]. Light pollution is expressed by the formula shown in **a** where $p_i$ and $ρ_i$ are the light pollution value and population proportion of a certain region in a country, respectively. After deleting the data points of eight outliers ( ● Cayman Islands, ● Iraq, ● Egypt, ● Israel, ● Moldova, ● Bhutan, ● North Korea, and ● Nepal), the trend in **a** became clearer. **b**[26]**|** More light is correlated with lower fertility. Although access to contraceptives has been the most important factor affecting fertility since the 1960s, light may also be an important factor. In the 1950s, the fertility (i.e., the number of live births per woman) in high-income countries was only 3, which may be related to their relatively high levels of nighttime light. In the 1950s, fertility started to drop in South Korea and Taiwan, coinciding with the increasing popularity of fluorescent lamps (Fig. 3), which exhibited 10 times higher illumination and 58 times more non-visual effects than the incandescent light bulb. In the same period of time, increasing utilization of TV also added too much light. Furthermore, children in these areas studied longer at night.

### Inspiration for improving human fertility from pig farming

If modern society intended to increase the fertility by scientific means, any big pig farm could provide reference, especially about how to use an artificially extended photoperiod. An ovulation rate of 20 is not uncommon in modern highly prolific sows. If all parameters (including the photoperiod) are controlled, a sow would have the potential to farrow 2.6 times/yr and to produce 52 pigs/yr[18], while the boar sow ratio is as high as one boar per 20 sows[19]. A boar has a tremendous influence on a farm's productivity and profitability[19]. Although a photoperiod of (15 h/d) (hours/day) could accelerate the boar maturation from 32 to 26 weeks of age[20,21], no one use this accelerating method. Boars must be carefully trained and not overused or abused even after 32 weeks of age. Their best working period only lasts 18 months[19]. For lactating sows, a supplemental light (16 h/d) in farrowing rooms could increase the weight of litters and result in a faster return to estrus after weaning[22,23]. For boars, an extended photoperiod could not only result in a faster sexual maturation but also a higher libido[21,24]. Furthermore, breeding sow estrus could be extended and occur more rapidly[9]. The lighting level and photoperiod for estrus listed in the ASABE Standard is only 100 Lx at 14–16 h/d[9]. For modern people, this value is at least 300 Lx at 16 h/d. Light stimulus occurs through the eyes, and clothes provide no protection.

Based on the above considerations, we could draw the following conclusions: 1) Light could accelerate boar maturation from 32 to 26 weeks. Applying the corresponding accelerating factor to predict the age of sexual maturation of men nowadays based on the average maturation age of 18 of men living 100 years ago gives an expected sexual maturation age of 18 × 26/32 = 14.3, which is in good agreement with Fig.3; 2) The questions arise if modern humans with extended photoperiod (16 h/d) are in a state of sexual excitement compared with previous generations living 70 years ago, and if sexual excitement is the essence of individual freedom or sexual liberation; 3) the reproductive ability of pigs has dramatically increased over the past 50 years. Therefore, the decline in human fertility is not primarily due to chemical pollution. Common sperm problems of modern men[25] may be mainly caused by premature development or over-frequent sexual activities stimulated by excessive light at night. The practice of weeding out boars with bad sex habits (e.g., masturbation)[19] on pig farms suggests that human beings should attach great importance to the reproductive health of adolescents; 4) blue light at light may be a hormone drug, a (sexual) stimulant, or even a drug( E2 in Fig.1) .

### Importance of the balance between light and dark, as well as the origin of consciousness and maternal love

Darkness on earth provokes a competition of all living things for sunlight. The abundance of air on earth prohibits any competition for air. If the availability of light was unlimited on earth, there would be neither competition nor evolution on earth. In other words, none of the currently living organism would exist**.** Therefore, our sense of competition, i.e., our desire and consciousness, may originate in the dark. Even men and women, the maternal love of females, and the territorial sense and chivalrous spirit of males are rooted in the dark, as they are all the results of competition. Therefore, we could philosophically say that light is the source of excitement, and dark is the source of desire. if the earth loses the dark, people will lose their desire.

The importance of maintaining a balance between dark and light is genetically predetermined. Only by maintaining this balance, we can normalize our mindset and consciousness, as well as our values and decisions. However, due to the desire for more light and longer hours of light exposure at night (i.e., blue is a stimulant), this balance was broken sharply by all kinds of electric light. Therefore, human beings should overall rethink decades of values and decision making, as well as reflect on behavior and moral standards. Perhaps it should be acknowledged that humans are hardly able to take rational decisions while being in a state of madness. The main issue is not only the sexual excitement of the individual by blue light at night; if we continue to lose this balance, we may not be able to save ourselves, e.g., with fertility problems described in the following and shown in Fig. 4 ( E4 in Fig.1).

### Too much light at night can lead to low fertility

Ref. 28 reviews 22 important factors that influence the fertility in the developed countries where contraceptives are readily available. These factors include micro, macro, and mesoscopic factors. However, no factor related to light is mentioned. we will not only fill this gap but also link all 22 factors to light. (see Supplementary Information for detail).

**Micro-Level**. It is the animal instinct of normal men and women to give birth more often, which is genetically determined. However, an extended photoperiod and high intensity light in childhood may avoid the full development of their reproductive system, which is being abused instead (as observed for boars[19], see above). So, today's youth may not develop to men or women as 70 years ago; the light at night has neutralized them and reduced their fertility.

On the other hand, night light also causes sexual excitation. People pursue individual freedom, but in essence, multiple sexual partners have showed up, which leads to various problems affecting fertility. Even more worrisome is that, if blue light is a sexual stimulant or drug, its control and harm to humans have the same characteristics as other stimulants, such as increasing dosage, addiction, no interest in other things, naturally no interest in raising children, and ultimately low desire.

As mentioned above, woman's motherly love and man's responsibility originated from the pursuit of light in the dark. Excessive illumination made men and women at the micro level only pursue shallow sexual excitement. Human beings 70 years ago were completely different both physically and mentally. Therefore, modern men and women who have become neutral cannot be expected to take the initiative to overcome all difficulties and give birth.

In modern society, people are under great pressure, which results from the comparison between people. If bearing more children is human instinct in the whole society, there will be no such pressure in life. Compared with human beings who lived hundred years ago, the life of modern people in advanced countries is very convenient. They have enough food and do not have to cut wood or wash clothes. Technological advances have made human life very convenient. Therefore, by affecting hormone secretion, it may be the blue light at night that led modern people into a state of madness.

**Macro-Level**. Until today, modern humans still hate the dark and consider the pursuit of light to be a noble thing, not realizing that they are in a deeply abnormal state. Policy makers, who have been also affected by light, only strive to satisfy the sensory stimulation of their voters and do not dare to take the necessary responsibility, for example, to control their voters' "inalienable right to sex freedom": they do not dare to ban contraception and abortion or unnecessary cesarean sections. Excessive light neutralizes modern people and makes them lose the natural maternal love and chivalry needed to maintain social civilization and achieve progress.

### Are low-desire societies derived from excessive light? (E5 in Fig.1)

China has 200 million unmarried young people older than 30 whose parents get together an anxious search for a mate for their son or daughter, forming spontaneous matchmaking corners in parks. While wild animals fight and risk their lives for the right to mate, the modern youth seem to generally have a low appetite. Moreover, it is difficult to hear unforgettable first love stories today. These observations remind of the terrible results of two free-breeding experiments on rodents conducted by the American ecologist John B. Calhoun in the last century[29–31]. When the rodents had enough food and drink, and there was no competition for nesting, no natural enemies, and no obvious infection—thus conditions that were realized in the rat utopia experiment—no new mouse was born after the rodent density rose to a certain extent, and the whole population withered. In the course of the experiment, young mice became neuter, they built no nests, did not fight and mate, and only quietly combed their hair. In other words, young mice had no desire. Furthermore, after these "beautiful rodents" were taken out to join a natural group, they still did not mate, which indicates that the process toward the development of low desire is irreversible. In Japan, South Korea, and China, people are worrying about whether low fertility is an irreversible result of a low-desire society.

Calhoun's results may be explained by the exposure to too much light. In his study, the second experiment was recorded over one million times while the lamp might be turned on. As rats are rodents and more sensitive to light intensity and photoperiod, they were inadvertently exposed to too much light.

Today, young people with smartphone addiction are likely to enter some state of low desire. If this phenomenon is irreversible, the society should immediately attach great importance to it. This is also consistent with our previous hypothesis of "dark is the source of desire; no dark, no desire".

It is worth noting that Calhoun's mice were inbred BALB/c mice, which do not degrade due to inbreeding, and BALB/c test mice are still sold today. Just as the reproductive capacity of pigs did not decrease due to environmental and food factors, BALB/c mice were also not evidently affected by peripheral chemical pollution or hormones. Therefore, light may be the main reason for the low desire of animals and even humans.

We also reached this conclusion (**M28**) based on the reproductive observation of mice in the laboratory (but no control group, E6 in Fig.1).

### The blind spot of human civilization in the digital era.

**Light and civilization.** Western civilization has brought a lot of convenience, abundant food, good medical treatment, and a longer life span to mankind. However, the fertility rate in the developed countries is low, and the Western civilization may be in danger. In China, the succession of dynasties often saw civilization overthrown by barbarism (during evolution, it is difficult to distinguish between civilization and barbarism). The theory of this article explains why the offspring of the ruling class became neuter to some extent and is no longer aggressive and enterprising by the exposure to too much light during receiving more education at night, which changed their hormone secretion. By classifying these conditions as "too much light", moonlight (0.2 Lx) is taken as a reference. A further result of this increased exposure to light is an Indulgent but impaired sexual life.

Today, everyone is praising the advantages of digital era, but it is neglected that the popularization of digital devices will result in an increased exposure to light at night. Despite its convenience, the popularization should be suspected to be a death blow for fertility, especially in the countries that first launched a new digital device. Therefore, in this "exciting" digital era, governments of all countries should work together to avoid this common social problem in advance.

At night, humans are psychologically seeking light, while physiologically no light should be available. Given our love of light and the fact that the intensity of daylight is sometimes as high as $10^5$ Lx, it is hard to believe that the power of 0.2 Lx at night can affect our reproductive system and behavior. In addition, human beings are adapted to the natural light cycle of 24 h (T = 24 h), but now people receive very frequent artificial light pluses at night (T ≈ 0.1–1 h). These light pulses disrupt the internal rhythm of human beings many times, leading to general inner restlessness and the feeling that time passes too fast. Too much and too mussy nightlight may make people unhappy and feel very busy and anxious.

### Conclusions and outlook on options to improve human fertility

Based on the experiments with hens, pigs, and mice, as well as the relation between origin of menstruation, sexual precocity, light pollution, and fertility rate, it has been suggested that the balance between light and dark is the foundation of this world and even the origin of human consciousness, which determines human values and behaviors. The significant change of the light environment during the last 70 years has thoroughly broken this balance and caused universal precocity and a general state of sexual excitement for modern people. However, it has not been realized that people are in a state of madness and that modern civilization is in danger, as not only the fertility rate is declining but also the human values and behaviors have become abnormal. Regarding the fast development of digital era, great importance must be attached to the negative impact of light on human civilization.

To increase fertility, people and policy makers first need to reach a normal state. As our consciousness and values are likely to be influenced by light, our values are not the same as those of people who lived 70 years ago. A typical example is how we look at the legalization of homosexuality. Identifying with the legitimacy of homosexuality may indicate that our hormone levels determine that we are already homosexual to some extent. Accordingly, identifying with young people with multiple partners before marriage in the name of personal freedom may indicate that we are also in a sexually excited state. Women should reflect whether they have enough maternal love to be good mothers to have more children or grandmothers to help with childcare. Men should consider if they have enough hormones or chivalry to keep society and their families in a normal state. Policy makers should reflect if their efforts are sufficient to lead a true civilization. In short, we cannot expect a modern human, who is in a state of madness, to take social responsibility and have more children. As a result, any policy that encouraged childbearing had ultimately been ineffective. Therefore, it is important to make humans aware by performing more meaningful animal experiments, such as performing Calhoun's experiment by adding the factor light to show that too much light leads to a low-desire society. Regarding fertility technology, pig-raising technology may be taken as a reference. We believe that controlling human exposure to light from a young age on may generally solve the problem of male and female fertility degradation.

Too much artificial light is endangering the populations of Europe, USA, China, Japan, and Korea, sparking clashes among civilizations. If there were enough Caucasian in the West, Europe would not have an immigration problem, nor would the West shift its production capacity to China, which would inevitably lead to "copy". So the unrest of the world could be rooted in "artificial light".

**Supplementary Information** is available in the online version of the paper.

**Acknowledgements.** This study was funded by the National Key R&D Program of China (No. 2017YFB0403700), National Natural Science Foundation of China (No. 61864008), and the program "Special scientific research project of Shaanxi Provincial Department of Education（20JK0615）".

**Author Contributions.** F.WQ and F.KQ conceived the study; F.WQ, Y.CP, and F.TQ designed and prepared experiments on hens and interpreted the data; F.WQ, Y.CP, C.TY, Z.YH, H.SX ,F.TD, Y.B, W.XD, and L.C carried out the experiments and records; F.WQ and F.KQ wrote the manuscript; and F.WQ, F.KQ, and Y.CP modified it.

**Author Information.** The authors declare no competing financial interests. Readers are welcome to comment on the online version of the paper. Correspondence and requests for materials should be addressed to F. WQ (fwq@ncu.edu.cn) or Y. CP (cpyang2022@sinano.ac.cn).

# Methods

**METHODS**

No statistical methods were used to predetermine sample size. The experiments were not randomized, and the investigators were not blinded to allocation during experiments and outcome assessment.

**M1:** We chose a research subject that is urgent, of common human concern, and has not been addressed to date. As mankind cannot directly be used as a sample to study the increase of fertility, we applied a method that uses association and analogy and is based on optics, evolutionary theory, hormonology, and even philosophy to analyze the essence of low fertility in the modern society from complicated reproductive phenomena and numerous related literatures. Low fertility may result from hormone changes induced by an increasingly extended photoperiod and increasingly higher light intensity at night, which not only affect the fertility but also the people's values. With the advent of the 5G era, the adverse effects of increased exposure to light on civilization need to be considered in a historical and philosophical context.

**M2: Why does light (photoperiod, spectrum, intensity, and stability) mainly affect the human reproduction system instead of food, chemical pollution, or social and psychological factors?** We addressed this question in the following six ways: 1) Division into endogenous and exogenous hormones; 2) over the past 100 years, the reproductive ability of pigs has increased; 3) over the past 70 years, the inbred reproductive ability of BALB/c mice has not declined; 4) in the past 100 years, the reproductive ability of dogs, cats, and monkeys did not decline; 5) our hen experiment showed that the eating start time has no influence on hen ovulation (ED Fig. 5), but light at a low intensity of 0.2 Lx was sufficiently powerful to affect hen ovulation. This conclusion was partly supported by the "sound or dark" experiment (ED Figs. 5 and 1, respectively); 6) social and psychological factors could also be attributed to the effects of light (see Supplementary Information)

**M3: Quantitative tests based on human samples have been performed**. After excluding dietary and chemical contamination factors in **M2** (the most distinctive step) and scaling up to 70–100 years, we can conclude that all of humanity is involved in the quantitative experiment on the relation between light and reproduction. Based on the definition of the illuminance unit Lx, the illumination value over 100 years can be roughly quantified, and the evolution of the age at menarche can be quantitatively related to the illumination. Based on NASA's population data for all countries' subregions that are affected by light pollution, as well as birth rates recorded by the United Nations over 70 years, it is possible to correlate current birth rates with photoperiod and night brightness. Therefore, the conclusions of this article are also based on two quantitative tests of human samples, and a large-scale human sample has been used for a long-time study of the fertility under artificial light.

**M4: Every single piece of evidence alone is not convincing, but the combined evidence indicates a significant effect of light on fertility.** We strive to simplify complex issues. For example, the strategies applied in pig industry to increase piglet rates has important implications for increasing human fertility. Although our approach is different from research at the molecular or cell level, it gives an explanation of the causes of low fertility (see Supplementary Information). We also designed animal experiments to demonstrate the effect of low light intensity, and the corresponding methods are described in the following.

**M5: Purpose, principle, and design of the lighting test of hens**. In average, a hen lays one egg every morning. In the new layer, which was kept at the constant temperature of 25 °C, hens could lay eggs for at least one consecutive year, which was sufficient to guarantee the stability of our experiment. In general, a hen lays about 1000 eggs in her lifetime[33], while a woman ovulates about 400 times in her lifetime[34]. Many similarities are observed between a hen laying eggs and human ovulation in terms of evolutionary history or genetics. However, there is a significant difference in the circadian rhythm of laying eggs and the monthly rhythm of ovulation, although both rhythms should be controlled by light.

**M6:** Two flocks of hens were used in parallel to ensure the reliability of the test results. The growth environment, dietary conditions, as well as disease and epidemic prevention of the two flocks of hens were strictly the same, but the illumination of the two flocks was reversed to verify the test results by referring to each other.

**M7:** We designed a novel method for counting eggs using an infrared video, which does not interfere with the rhythm of the hens (see Extended Data). This method uses egg laying as an indicator of the hen's circadian rhythm instead of taking blood samples or measuring the temperature, which facilitates the experiment and does not cause pain to the hen. We applied various test conditions and measures, which are described below.

**M8:** Infrared cameras were used to count eggs in only ten seconds while eliminating disturbance from the camera. In each of hen house A (Flock A) and hen house B (Flock B) three groups of hens were housed (ED Fig. 7b). For each group, we set up a video camera. The video head was equipped with an 850 nm infrared LED light source to facilitate observing hens and eggs in the dark. Videos were recorded for 10 s every 1.5 h. Then, the power of the camera was cut off to avoid that the tiny red light of the 850 nm LED light source affected the hens. This measure was important, as hens are especially sensitive to red light (ED Fig. 5). Then, we videotaped the eggs outside the hen house to obtain the total number of eggs laid in 1.5 h.

**M9:** The following light parameters were applied: lighting was divided into feeding light (ED Fig. 8b) and treatment lights (ED Fig. 7a and ED Table 1). Hens have four types of photoreceptors[35], and in most of the reported hen experiments, light parameters, such as the color rendering index, flicker, color temperature, etc., were not sufficiently controlled. To keep the light as close to natural light as possible, we strictly defined the parameters of the light source: dietary (food) illumination was 50 ± 5 Lx; the lamps were self-made; the phosphors were excited by a 450 nm blue LED; the color temperature was about 4500 K, the color rendering index was about 90, and there was no strobe. If not indicated otherwise, the treatment light was emitted by a purple LED of the self-made lamp. This lamp was a DC dimming lamp with full visible spectrum emission, and no PWM dimming was used to strictly avoid any flicker. Moreover, to further mimic natural light, the lamp was also supplemented with an 850 nm infrared LED light source. The color temperature of the treatment light was 2700 ± 100 K with Ra ≥ 93. The Lx meter used was TES1330A DIGITAL Lux METER.

**M10: Identical eating time and light environment during eating.** For both flocks of hens, the feeding times were 6:45–8:15 and 18:45–20:45, and the illumination near the hens' eyes was set to 5 Lx (15 min) + 50 Lx (60 min) + 5 Lx (15 min). In the 15 min before and after feeding a buffer of light was applied to prevent the hens from being frightened. The hens were fed with "laying hen feed" produced by Haida Company. If the hens continued to receive treatment light after feeding was completed, the feeding trough (ED Fig. 7b) was removed artificially. If the hens should enter the dark environment, the feeding trough was not removed. Moreover, the hen's automatic water fountain was never removed.

**M11: The hen coop was relatively comfortable.** An industrial hen coop (battery cage system) with egg-rolling-out function was adopted.

# Methods

Each coop was built to house four hens, but we only housed three hens per coop for their comfort (ED Fig. 7b). Each hen coop was equipped with an automatic drinking fountain, and water was obtained from the city water supply. The battery cage system was divided into three layers with uniform illumination at each layer. We removed hen droppings every afternoon during feeding time.

**M12: Purification by air conditioning at a constant temperature.** The growth and development of hens are greatly affected by the temperature, which indicates that the temperature factor needs to be excluded in this test. The hen houses were built by transforming an old air conditioning purification laboratory (ED Fig. 7b). The two hen houses shared a circulating air conditioning unit with a constant temperature of $25 \pm 0.5$ °C. This system used a medium efficiency particulate air filter and was equipped with an ultraviolet lamp in the air conditioning cabinet to eliminate viruses. The coarse filter was cleaned every day, and the bag filter in the air conditioning cabinet was cleaned every week to ensure a good air quality in the hen house. White color steel sandwich panels served as the wall plate and caused the hen coop to shine evenly. An air supply fan was controlled by a variable-frequency governor to avoid any obvious wind sensation at the hen's position. Fresh air supply for the hens was ensured by 24 h exhaust and fresh air exchange.

**M13: Breed and source of laying hens.** In total, 208 hens were purchased from an industrial hen farm (Fuzhou Fuxin Industrial Co., LTD.), and the name of the breed was "Jingfen". When the hens were delivered (2019-7-18), the age of the hens was 60 days. The hens began to lay eggs at 120 days of age. When the hens were 150 days old and egg laying was stable, the formal lighting experiments were started.

**M14: Immunization of laying hens:** Laying hens was immunized four times according to the experimental plan. To avoid scaring the hens and reducing their emergency response, we gave the injections in the dark, so no influence on egg laying was detected. Only one injection was completed at the age of 150 days, and the test was delayed by five days after this last injection. We were especially concerned about the health of the hens and were able to ensure good health throughout all tests.

**M15: Stray light and noise control in the test environment:** With double walls, double doors, aluminum foil tape, and opaque curtains, we turned the hen house into a dark room. We separated treatment light from feeding light. When the treatment light of hen House A was turned on, the test light of hen house B is turned off and vice versa. The sound insulation between hen houses A and B was also good. The noise outside the laboratory had little impact on the hens.

**M16: Standard illumination test with a light intensity of 10 Lx.** A light intensity of 10 Lx (ED Fig. 8a) was used to initialize the hens by bringing all hens into an equal state before any other light test was performed. After the hens were purchased, test light conditions of 10 Lx were selected, and the results showed that the egg-laying phases in hen houses A and B were indeed opposite. About five days after reversing the light, the egg-laying phases were significantly inverted. This process was repeated twice and proved that light at the intensity of 10 Lx could effectively reverse the egg-laying phase of the hens. It also proved that this method of egg counting was a very effective and convenient way to study the circadian rhythm.

**M17: Dark experiment (ED Fig. 1).** If the egg-laying rhythm every day is controlled by light, then the egg-laying phase should remain unchanged in the dark. These expectations have been confirmed by the test results. We performed experiments in the dark over a total of 20 days without detecting an evident change of the egg-laying phase.

**M18: Lighting test at an intensity of 0.2 Lx (Fig. 1).** Considering that the egg-laying phase of a hen does not change in the dark, we investigated in this experiment how much light can change the egg-laying phase of a hen. As the light intensity, we chose the ground illumination of the full moon of 0.2 Lx (ED Fig. 8a). At this illumination, the egg-laying phase of the hens could be reversed within nine days.

**M19: Comparison of the effect of illumination at 0.2, 3, and 10 Lx (ED Figs. 4 and 8a).** In this experiment, we studied the egg-laying phase under illumination with different light intensities (0.2, 3, and 10 Lx) to assess if the effect of illumination has a saturation value. The experiment showed that the light intensity of 10 Lx was close to the saturation value. Under illumination at 10 Lx, the egg-laying phase changed about three times faster than under illumination at 0.2 Lx.

**M20: Comparison between the effects of brightness and illumination (ED Figs. 6 and 7a).** Although the illumination of the moonlight on the ground has an intensity of only 0.2 Lx, the surface brightness of the full moon is relatively high. A legend in China says that "rabbits get pregnant when looking at the moon", which raises the questions whether the brightness of the moonlight has a major influence on the circadian rhythm and whether animals receive a signal from the moon that controls their rhythm. Therefore, we designed and performed different experiments at a constant illumination of 0.2 Lx. In the first test, the hens could see the LED beads, and the brightness of the beads was very high and even dazzling. In the next test, the hens could not directly look at the lamp bead, and the illumination of 0.2 Lx was achieved by diffuse reflection. The results showed that under a constant illumination of 0.2 Lx near the eyes of the hens, no difference in the effects between brightness and illumination was observed.

**M21: Influence of illumination on hens exposed to three shifts lighting (ED Fig. 2).** In this experiment, we divided the 24 h of a day into three intervals of 8 h and inverted the lighting (10 Lx) after every interval of 8 h to expose hens to three daily lighting intervals instead of the two daily lighting intervals of day and night. While humans working in three shifts face difficulties falling asleep in time, we noticed that laying hens could be quiet within 1 h in the dark, and exposure to three daily lighting intervals had no significant influence on the total number of eggs laid by the whole flock, so one egg was still laid in 24 h. However, the egg-laying phase of the hens could be disordered, and the egg-laying peak was less obvious.

**M22: Comparison among the effects of illumination with three RGB wavelengths (ED Figs. 5 and 8d).** Under the condition that the number of photons was roughly constant, the influence of red LED (630 nm, 1.3 Lx), green LED (525 nm, 6.0 Lx), and blue LED (460 nm, 0.22 Lx) on the egg-laying behavior was studied. The results showed that blue light had the smallest influence on the egg-laying behavior, which may be related to visual and non-visual effects on the hen, as red light is known to exert more non-visual effects through the hen skull[36].

**M23: Effect of light without blue component (ED Figs. 3 and 8c).** The influence of a combination of yellow and red light on the egg-laying phase was studied in the absence of blue light, showing that light without blue component still had a significant influence on the egg-laying phase. The most sensitive wavelength of humans is quite different from that of hens.

**M24: Influence of the eating start time on the egg-laying phase (ED Fig. 5).** As mentioned earlier, hens usually lay eggs in the morning, which raises the question if the egg-laying phase depends on the sunrise or on the feeding time in the morning. Our results showed that egg laying depended on the sunrise. After the eating time was shifted by 6 h from (6:45–8:15) and (18:45–20:15) to (12:45–14:15) and (0:45–2:45), the maximum egg production did not shift by 6 h, as would have been expected if egg laying depended on the eating time. Therefore, light is much more important than food for the reproduction

## Methods

of both humans and hens when food is abundantly available, which also indicates that chemical pollution is not as harmful as light pollution.

**M25: Influence of sound in the dark (ED Fig. 5)**. We tried to influence the egg-laying phase of the hens by sound instead of light. The sound was turned up to an estimated volume of 70 dB, and Buddhist music was played. However, the hens showed no reaction to the sound, and the hens' sleep was unaffected by the sound. Furthermore, no change in the original egg-laying phase was observed.

**M26: Calculation of the data points in Fig. 4a**. The light pollution value $P_{sum}$ (mcd/m$^2$) was calculated according to the ratio of the weight of the population. In the literature[27], light pollution was divided into six stages, namely $P_1$–$P_6$. By using $\rho_i$ to represent the proportion of the population affected by $p_i$, the following values were obtained for the case of the United Kingdom[27]:

$P_1 < 1.7$, $\rho_1 = 0\%$; $1.7 \leq P_2 < 14$, $\rho_2 = 0.1\%$; $14 \leq P_3 < 87$, $\rho_3 = 1.7\%$; $114 \leq P_4 < 87$, $\rho_4 = 1.7\%$; $87 \leq P_5 < 680$, $\rho_5 = 21.2\%$; $680 \leq P_6 < 3000$, $\rho_6 = 51.0\%$;

$P_{sum} = \rho_1 LgP_1 + \rho_2 LgP_2 + \rho_3 LgP_3 + \rho_4 LgP_4 + \rho_5 LgP_5 + \rho_6 LgP_6 = 0\% \times Lg1.7 + 0.1\% \times lg14 + 1.7\% \times Lg87 + 21.2\% \times Lg680 + 51.0\% \times Lg3000 = 2.78$

and the total fertility in the UK[26] was 1.75, which explains the data point (2.78, 1.75) in Fig. 4a. As human perception of light is logarithmic, Log Pi is plotted in Fig. 4a.

**M27**. **By demonstrating that menstruation originates from the change of moonlight over a month, we confirmed that the human reproductive system is also very sensitive to light at a low intensity**. The menstruation of all primates is not strictly synchronized with the moonlight, as the moonlight is too weak, susceptible to the weather, and the amount of moonlight every individual receives is very different. Young people in previous generations were more energetic and received more moonlight for a longer time outside. It is also this mismatch that suggests that the human reproductive system is sensitive to dim light. We applied deductive reasoning using a method of elimination based on the following statements: 1) In primates, menstruation is regulated by circalunar clocks, similar to non-biting midges[37], as it is determined by the genes; 2) the circalunar clock is not regulated by the tides, and there is no relation with the universal gravitation of moon and sun. The relationship between the decline in sea turtles and light pollution has shown that light was the decisive factor[38]; 3) the human reproductive system is independent on the temperature; 4) the human reproductive system is independent on food as long as food is abundantly available; 5) the human reproductive system is independent on sunlight, which affects the circadian rhythm; 6) there is no need for humans to evolve any additional photo rhythm sensor except for that detecting the non-visual effects of light (ipRGC); 7) only primates are characterized as forest-dwelling, social animals whose activity at night is determined by the moonlight, and their eyes are equipped with rod cells that sense dim moonlight; 8) taking monkeys as an example, our field study showed that the Monkey King is usually stronger and more intelligent than the other monkeys. He keeps a firm grip on six or seven concubines during the day to drive off competitors, who only have chance at full moon; 9) due to their smaller weight and higher activity and liveliness, little monkeys could climb up to higher places where they were exposed to more moonlight. After sexual maturation, moonlight helped them mate with stronger young males from outside the group. Ovulation at this time, as long as no inbreeding occurred, presented an evolutionary advantage for the species, which gradually resulted in the evolution of a menstrual phenomenon; 10) older females, unwilling to move into the moonlight at night, gradually lost their menses synchronization[39,40] with the moonlight. This disorder was advantageous for the Monkey King to rest, which was also better for the competition of the species.

**M28. The reproductive observation of mice in the laboratory in our Univ**. Our colleagues inadvertently kept the mice in a room that lamps were turn on from 9:am to 9:pm automatically. Since there were no curtains in the room, the total light may last 15h a day. It was common that the mice, after a few generations, had significantly reduced fertility.

**M29. Influence of mechanical disturbance.** We started two rounds of laying hen experiments in 2019 and 2023, with the second round complementing the first round of mechanical disturbance trials that were not completed due to the COVID-19 pandemic. The purpose is to prove that the laying phase of laying hens cannot be changed by reversing the day-night phase of mechanical disturbance in the dark. This is indeed the case, suggesting that children who listen to learning in the dark at night do not interfere with hormone production to some extent.

# Extended Data

## Our egg-counting method may be used to study the effects of light on the ovulation phase

**Abstract:** The Extended Data mainly resulted from eight ovulation experiments where our egg-counting method applied (see **Methods**). All conclusions presented below may be related to humans, and the detailed description and conclusions of the hen experiments are described in the following. **"Experiments"** of the effects of light on pigs, rats, and humans are described in the main text. Here, the term **"Experiments"** indicates only those "hypothetical experiments" that were performed on a super large scale over a very long time, and were conducted by other groups or by Nature (God), which used pigs, mice, and even humans as samples. All types of experiments confirmed our hypothesis, which was also corroborated by the analysis of human samples shown in the Supplementary Information. **We especially focused on demonstrating that chemical contamination of food is not the main cause of the decline in fertility.**

| | |
|---|---|
| Zero Light or Dark | **Extended Data (ED) Fig. 1**. **Results**: Zero light or darkness has no effect on the egg-laying phase of hens. **Hints**: The general expectation to get energy from sunlight causes problems. For ovulation, light pollution is much more harmful than chemical pollution of food. |
| Egg Laying in Three Shifts | **ED Fig. 2. Results**: Laying eggs in three shifts has clear effects on the ovulation phase of hens. **Hints**: A stable photoperiod is very important. Laying eggs in three shifts affects ovulation negatively, while it is not clear if this result is transferable to women at childbearing age. However, it is known that strobe light has a negative effect. |
| Zero Blue Light | **ED Fig. 3. Results**: Zero blue light has clear effects on the egg-laying phase of hens. **Hints**: The spectrum that affects the ovulation of hens is different from that of humans. Human (hormone) rhythm is more affected by blue light, and the rhythm of hens is more affected by red light. |
| Light Intensity | **ED Fig. 4. Results**: A light intensity as low as 0.2 Lx affects the egg-laying phase of hens. **Hints**: Full moonlight (0.2 Lx) and sunlight at noon ($10^5$ Lx) may have similar effects on the egg-laying phase. This means that the effect of the photoperiod is most important, followed by that of the light intensity with a probable saturation value of 10 Lx. |
| RGB Light and Spectrum | **ED Fig. 5. Results**: The effect of visible light of different wavelengths decreases in the order R > G > B. Red light has the largest effect on the egg-laying phase under the condition that the numbers of photons of the different wavelengths of RGB light are approximately equal. **Hints**: In addition to the photoperiod, intensity, and stability, the spectrum has an important effect. |
| Brightness vs Illumination | **ED Fig. 6. Results**: Brightness and illumination have comparable effects on the egg-laying phase of hens. **Hints**: Animal ovulation is not changed by detecting the brightness of the moon surface, but it is changed by detecting illumination. |
| Eating StartTime | **ED Fig. 7. Results**: The eating start time has no effect on the egg-laying phase of hens. **Hints**: With enough food or nutrition, light is the key factor to determine the ovulation, which also indicates that the influence of toxic chemicals in food or air is much less than that of light. |
| Sound or Music | **ED Fig. 8. Results**: The eating start time has no effect on the egg-laying phase of hens. **Hints**: It is impossible to change the circadian rhythm of hens by using sound. |

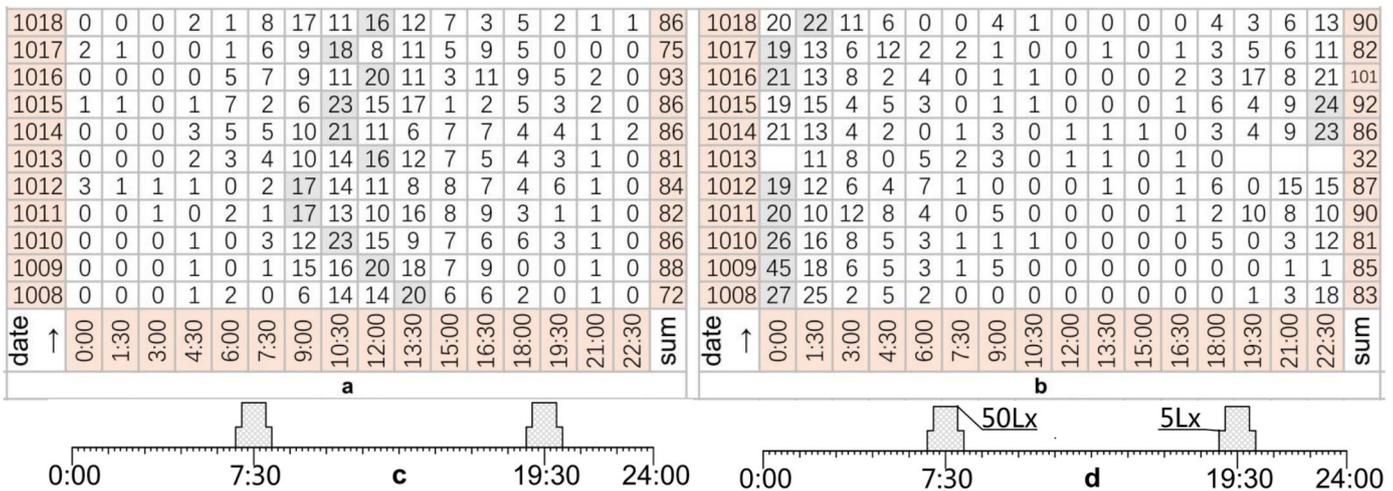

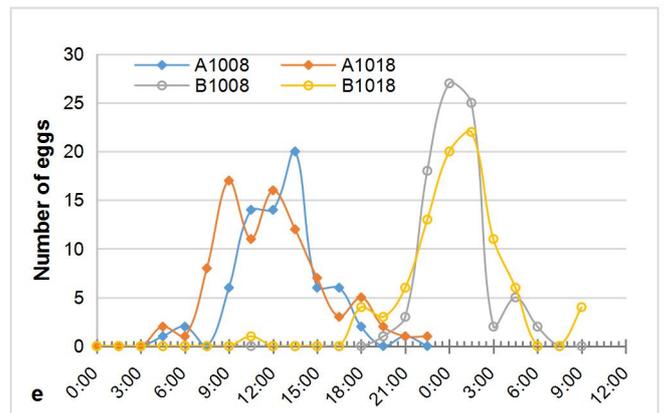

**Extended Data Fig. 1 | In the dark, the egg-laying phase of hens remained unchanged**, and no significant change was observed at least in 11 days from "1008" (2019-10-08) to "1018". **a |** The maximum egg production of Flock A maintained at around 10:30–12:00. **b |** The maximum egg production of Flock B maintained at around 22:30–24:00, which proved that light was the main factor that determined the egg-laying phase. **c, d |** Dietary lighting for hens (one lighting period in the morning and the other one in the evening) was composed of three different lighting intervals: 5 Lx (15 min) + 50 Lx (60 min) +5 Lx (15 min). **e |** Distribution diagram of the eggs laid in Flocks A and B in 24 h at the beginning (1008) and the end (1018) of the experiment. This graph further shows that the egg-laying phase is nearly constant in the dark. The "sum" of laid eggs in a and b shows no significant decline in the total number of eggs laid per day in the dark, which indicates that the energy that hens use to lay eggs is mainly derived from food and not from the sun or from any artificial light.

1Institute of material Science, Nanchang University, 235# NanJinDong Road, Nanchang, JIangxi 330046,P.R.China. 2 School of Chemistry and Materials, Shangluo College, Shangluo, Shaanxi，P.R.China,
†Present addresses:　fwq@ncu.edu.cn，cpyang2022@sinano.ac.cn

# Extended Data

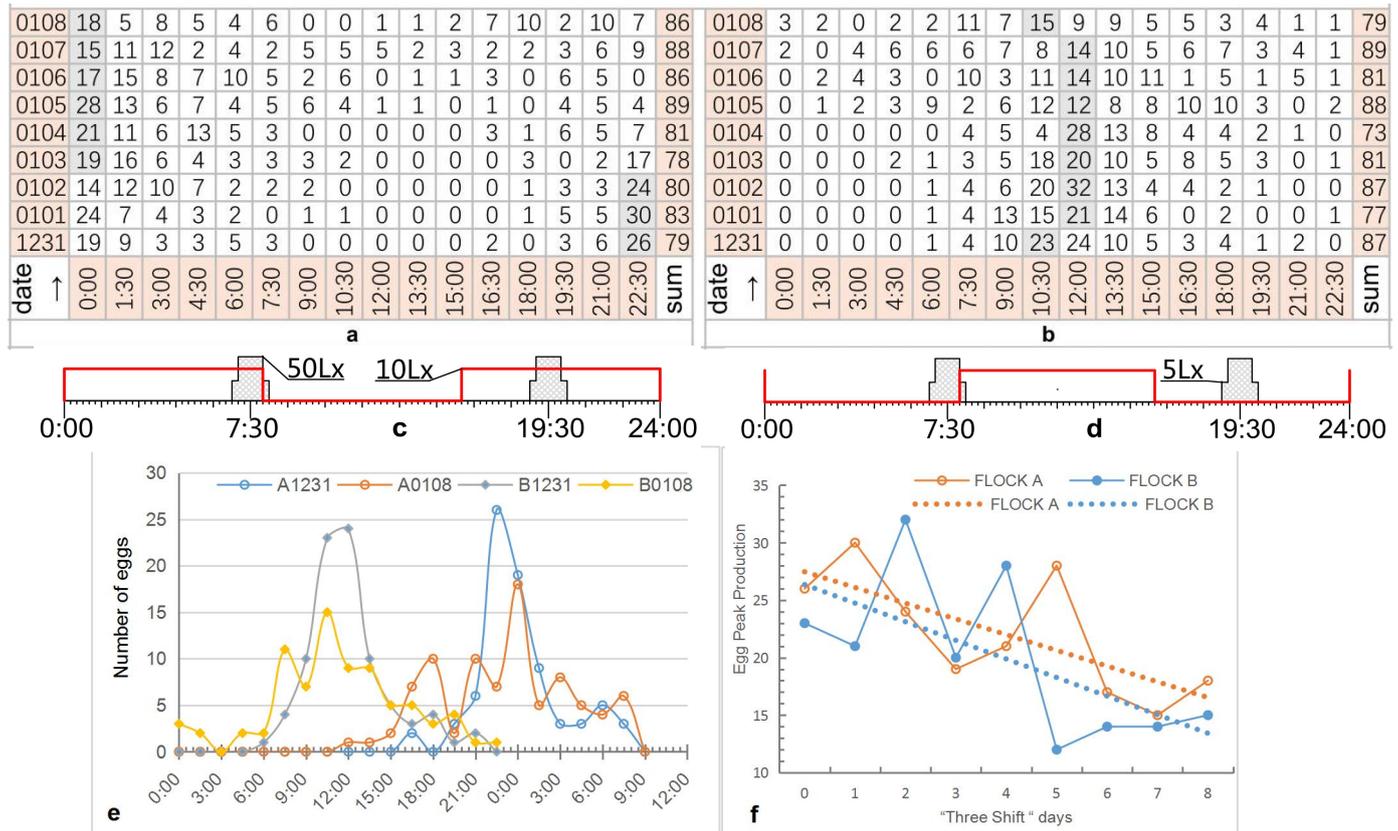

**Extened Data Fig. 2 | Importance of a stable photoperiod.** Exposure to a rhythm of three lighting intervals could disturb the egg-laying phase. What about women in three shifts? **a, b |** Hens in Flocks A and B were exposed to three daily lighting intervals starting on 2019-12-31 (1231). During the experiment, the peak position (phase) of laid eggs remained unaffected, but the peak decreased, and the distribution of eggs laid within 24 h became wider. **c, d |** The three lighting intervals in Flocks A and B, namely a sequence of 8 h of 10 Lx light and 8 h of darkness, were adapted to the daily 24-h rhythm (8 h × 3 = 24 h). Therefore, the initial peak of the egg-laying phase was barely maintained. **e |** Distribution of eggs laid at the beginning and the end of the test in Flocks A and B. For A0108/B0108, the peak of laid eggs was still visible, but the egg-laying phase of most hens has become disordered compared with A1231/B1231. **f |** With an increasing number of days where the hens were exposed to the rhythm of three lighting intervals, the peak gradually decreased. The dotted trend lines for Flocks A and B, as well as their slopes, were similar, which proved a downward trend. This trend line predicted that the egg peak production will disappear about 12 days after the beginning of the test, i.e., the exposure to three lighting intervals will seriously interfere with the reproduction of the hens. The "sum" of laid eggs in **a** and **b** showed that the total number of eggs laid per day has not been affected by exposure to three lighting intervals. This may be explained by the very short memory of the hens, which avoided that exposure to three lighting intervals had any psychological impact. However, this experiment shows that three daily lighting intervals will not only make people feel very uncomfortable due to psychological factors but also seriously interfere with their reproductive system. Therefore, human activities should be carefully adapted to the two daily lighting intervals of day and night.

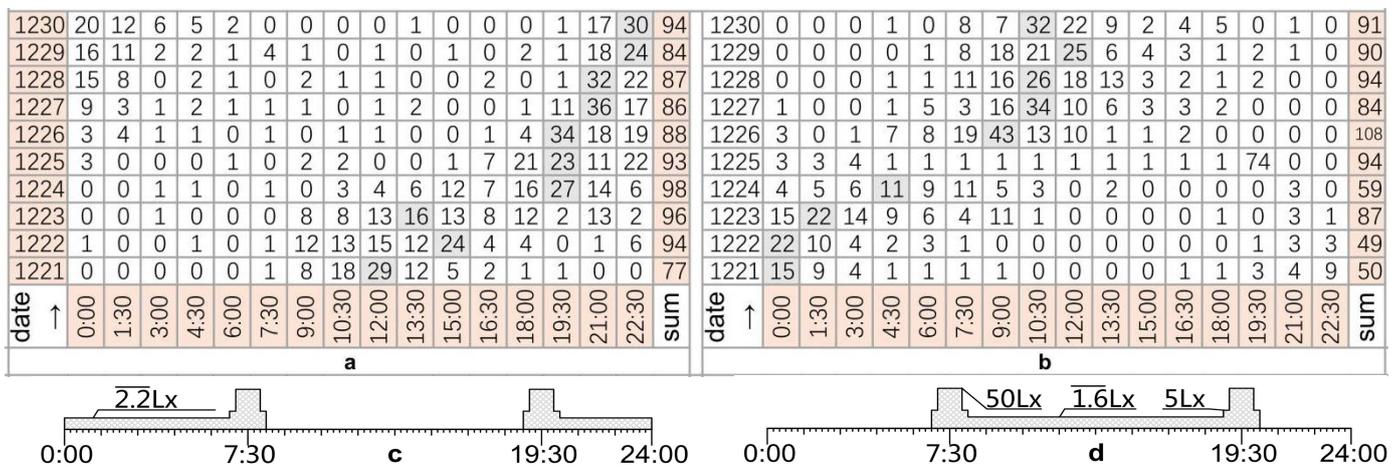

**Extended Data Fig. 3 | Zero blue light affects the egg-laying phase of hens.** The non-visual effects of light on humans are different from those on hens. The spectrum of our self-made zero-blue LED bulb is shown in ED Fig. 8c. **a, b |** Each of flocks A and B were equipped with a zero-blue LED bulb. Depending on the distance of the hens from the LED bulb (ED Fig. 7b), the illumination in hen house A was 3.2, 2.2, or 0.9 Lx, and the illumination in hen house B was 3.5, 1.6, or 0.6 Lx. **c, d |** The lighting phases of the two hen houses were exactly opposite to each other. Starting on 19-12-21 (1221), this light stimulated the hens to lay eggs in reverse rotation in about seven days. This experiment showed that blue light, which has an important effect on the human circadian rhythm, is not a key factor in hens.

# Extended Data

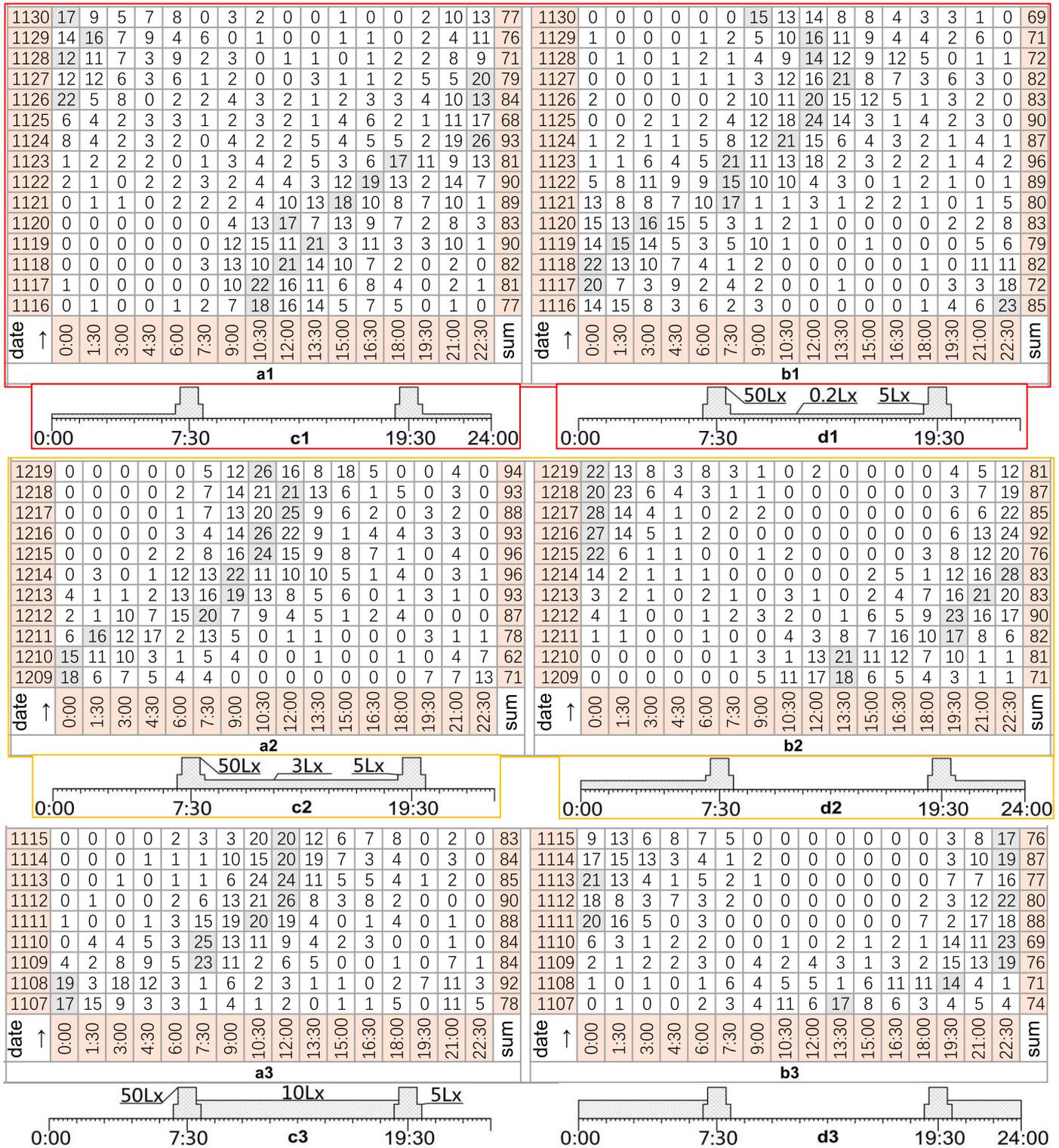

**Extended Data Fig. 4 | Higher illuminance is related to a faster change of the egg-laying phase, but it soon reaches a saturation value of around 10 Lx for diurnal animals, including humans**. **a1–d1 |** Under illumination at an intensity of 0.2 Lx, it took about nine days to complete the transition of the egg-laying phase. **a2–d2 |** Under illumination at an intensity of 3 Lx, it took about six days to complete the transition of the egg-laying phase. **a3–d3 |** Under illumination at an intensity of 10 Lx, it took in average about three days to complete the transition of the egg-laying phase.

If the period of the biological clock is T, it may take nT to reverse the egg-laying phase, where n ≥ 3, and n is determined by the light intensity. A light intensity of 10 Lx seems to be the saturation value for changing the phase of an animal rhythm. As a light intensity of 0.2 Lx can reverse the egg-laying phase of hens in nine days (n = 9), it may take 9 × 30 = 270 days to reverse the human ovulation phase at this light intensity (T = 30 days). For a light intensity of 10 Lx, it may still take 3 × 30 = 90 days. Based on these findings, we could give a new explanation why human menses today are not strictly synchronized with moonlight (see **M27** for details): humans had learned to use fire, and the firelight is both stronger and more irregular than moonlight; The process synchronizing human menses with moonlight in 90 days is too long and too complex to control.

# Extended Data

## a

| | date ↑ | 0:00 | 1:30 | 3:00 | 4:30 | 6:00 | 7:30 | 9:00 | 10:30 | 12:00 | 13:30 | 15:00 | 16:30 | 18:00 | 19:30 | 21:00 | 22:30 | sum |
|---|---|---|---|---|---|---|---|---|---|---|---|---|---|---|---|---|---|---|
| Flock B End Blue Test and Food Test Begin | 0217 | 8 | 19 | 20 | 9 | 15 | 8 | 1 | 0 | 2 | 6 | 1 | 1 | 0 | 0 | 0 | 1 | 91 |
| | 0216 | 21 | 9 | 5 | 8 | 9 | 8 | 0 | 0 | 0 | 2 | 0 | 0 | 0 | 0 | 1 | 3 | 66 |
| | 0215 | 23 | 18 | 4 | 7 | 5 | 5 | 4 | 1 | 1 | 3 | 0 | 1 | 0 | 4 | 7 | 17 | 100 |
| | 0214 | 27 | 11 | 13 | 5 | 1 | 3 | 1 | 2 | 1 | 0 | 0 | 2 | 1 | 0 | 2 | 9 | 78 |
| | 0213 | 18 | 16 | 7 | 0 | 4 | 3 | 1 | 2 | 1 | 0 | 0 | 0 | 2 | 8 | 14 | | 76 |
| | 0212 | 15 | 17 | 6 | 7 | 5 | 1 | 3 | 1 | 0 | 0 | 0 | 1 | 3 | 9 | 18 | | 86 |
| | 0211 | 13 | 16 | 10 | 3 | 6 | 3 | 1 | 1 | 0 | 1 | 0 | 1 | 0 | 3 | 0 | 17 | 75 |
| | 0210 | 10 | 26 | 10 | 7 | 5 | 3 | 2 | 0 | 0 | 1 | 0 | 0 | 5 | 11 | 13 | | 94 |
| | 0209 | 13 | 21 | 14 | 11 | 3 | 1 | 3 | 0 | 1 | 0 | 0 | 0 | 1 | 2 | 8 | | 78 |
| | 0208 | 31 | 16 | 4 | 3 | 0 | 2 | 1 | 0 | 0 | 0 | 0 | 0 | 3 | 2 | 8 | | 70 |
| Flock A End Red Test and Sound Test Begin | 0207 | 29 | 9 | 6 | 1 | 1 | 2 | 1 | 0 | 0 | 0 | 0 | 0 | 4 | 1 | 18 | | 72 |
| | 0206 | 32 | 4 | 2 | 3 | 0 | 0 | 4 | 0 | 0 | 0 | 0 | 0 | 2 | 0 | 27 | | 74 |
| | 0205 | 21 | 5 | 0 | 0 | 2 | 0 | 0 | 0 | 0 | 0 | 0 | 6 | 0 | 29 | | | 63 |
| | 0204 | 10 | 5 | 0 | 3 | 0 | 0 | 0 | 0 | 0 | 0 | 2 | 6 | 31 | 23 | | | 80 |
| | 0203 | 7 | 2 | 2 | 2 | 0 | 0 | 0 | 1 | 0 | 0 | 0 | 2 | 2 | 19 | 2 | 26 | 65 |
| | 0202 | 5 | 2 | 0 | 2 | 0 | 0 | 0 | 1 | 0 | 0 | 1 | 5 | 7 | 28 | 1 | 16 | 67 |
| | 0201 | 2 | 1 | 1 | 0 | 0 | 0 | 1 | 4 | 0 | 3 | 3 | 6 | 16 | 35 | 0 | 5 | 77 |
| Flock A Start Red Test, Flock B Start Blue Test | 0131 | 0 | 0 | 0 | 0 | 0 | 2 | 2 | 7 | 2 | 10 | 13 | 19 | 11 | 7 | 2 | 2 | 77 |
| | 0130 | 0 | 0 | 0 | 0 | 1 | 1 | 13 | 7 | 18 | 24 | 11 | 5 | 5 | 1 | 0 | 0 | 86 |
| | 0129 | 0 | 0 | 0 | 0 | 0 | 4 | 22 | 23 | 23 | 15 | 1 | 3 | 1 | 0 | 4 | 1 | 97 |
| | 0128 | 0 | 0 | 0 | 2 | 3 | 11 | 23 | 23 | 16 | 8 | 2 | 3 | 1 | 2 | 0 | 0 | 94 |
| Use 10Lx White Light to Consolidate | 0127 | 0 | 0 | 2 | 3 | 6 | 9 | 18 | 16 | 10 | 6 | 11 | 2 | 1 | 1 | 0 | 0 | 85 |
| | 0126 | 1 | 2 | 3 | 3 | 8 | 12 | 11 | 8 | 11 | 19 | 4 | 5 | 4 | 2 | 0 | 0 | 93 |
| Green Light Test End | 0125 | 0 | 2 | 3 | 3 | 7 | 5 | 11 | 6 | 13 | 19 | 8 | 2 | 3 | 0 | 1 | 0 | 83 |
| | 0124 | 3 | 1 | 3 | 6 | 6 | 6 | 9 | 11 | 16 | 13 | 4 | 3 | 4 | 0 | 2 | 0 | 87 |
| | 0123 | 2 | 2 | 0 | 4 | 5 | 4 | 13 | 17 | 14 | 11 | 8 | 3 | 2 | 0 | 0 | 0 | 85 |
| | 0122 | 1 | 5 | 4 | 1 | 6 | 10 | 11 | 14 | 26 | 4 | 2 | 5 | 6 | 0 | 0 | 1 | 96 |
| | 0121 | 1 | 4 | 1 | 5 | 6 | 12 | 17 | 14 | 11 | 8 | 4 | 5 | 0 | 0 | 0 | 1 | 89 |
| | 0120 | 1 | 4 | 3 | 3 | 12 | 14 | 9 | 18 | 14 | 6 | 0 | 0 | 3 | 1 | 0 | 3 | 91 |
| | 0119 | 3 | 3 | 3 | 22 | 10 | 9 | 22 | 7 | 4 | 5 | 2 | 1 | 0 | 0 | 0 | 1 | 92 |
| | 0118 | 8 | 9 | 14 | 8 | 13 | 13 | 10 | 7 | 2 | 0 | 3 | 2 | 1 | 0 | 1 | 2 | 93 |
| Flock A,B Start Green Light Test | 0117 | 8 | 18 | 11 | 13 | 16 | 4 | 8 | 2 | 3 | 1 | 0 | 0 | 0 | 0 | 5 | | 89 |
| | 0116 | 26 | 15 | 12 | 14 | 4 | 5 | 1 | 3 | 0 | 1 | 0 | 0 | 1 | 0 | 3 | 2 | 87 |
| | 0115 | 34 | 16 | 4 | 4 | 6 | 3 | 3 | 0 | 0 | 0 | 0 | 1 | 0 | 1 | 6 | | 78 |
| | 0114 | 17 | 20 | 7 | 6 | 3 | 1 | 1 | 1 | 0 | 0 | 0 | 0 | 1 | 3 | 14 | | 74 |
| | 0113 | 18 | 7 | 1 | 1 | 1 | 1 | 0 | 0 | 0 | 0 | 0 | 4 | 3 | 18 | | | 54 |

## b

| date ↑ | 0:00 | 1:30 | 3:00 | 4:30 | 6:00 | 7:30 | 9:00 | 10:30 | 12:00 | 13:30 | 15:00 | 16:30 | 18:00 | 19:30 | 21:00 | 22:30 | sum |
|---|---|---|---|---|---|---|---|---|---|---|---|---|---|---|---|---|---|
| 0218 | 0 | 0 | 0 | 0 | 3 | 7 | 13 | 13 | 31 | 6 | | | | | | | |
| 0217 | 0 | 0 | 0 | 0 | 4 | 6 | 18 | 11 | 23 | 5 | 8 | 2 | 3 | 2 | 1 | 0 | 83 |
| 0216 | 0 | 0 | 0 | 0 | 4 | 8 | 15 | 17 | 19 | 8 | 5 | 0 | 2 | 2 | 1 | 0 | 81 |
| 0215 | 0 | 0 | 0 | 0 | 9 | 9 | 12 | 21 | 19 | 8 | 6 | 1 | 1 | 2 | 2 | 0 | 90 |
| 0214 | 0 | 0 | 0 | 0 | 8 | 24 | 24 | 14 | 17 | 6 | 3 | 0 | 0 | 3 | 1 | 0 | 100 |
| 0213 | 0 | 0 | 0 | 0 | 10 | 19 | 24 | 14 | 17 | 6 | 3 | 0 | 0 | 0 | 0 | 0 | 93 |
| 0212 | 0 | 0 | 1 | 7 | 16 | 21 | 15 | 10 | 18 | 3 | 0 | 0 | 0 | 1 | 2 | 0 | 94 |
| 0211 | 2 | 2 | 9 | 10 | 10 | 19 | 8 | 14 | 10 | 4 | 4 | 2 | 0 | 1 | 0 | 0 | 95 |
| 0210 | 2 | 2 | 9 | 10 | 10 | 19 | 8 | 14 | 10 | 4 | 4 | 2 | 0 | 0 | 0 | 0 | 94 |
| 0209 | 5 | 12 | 18 | 7 | 5 | 6 | 19 | 2 | 9 | 6 | 0 | 3 | 2 | 0 | 0 | 0 | 94 |
| 0208 | 2 | 13 | 16 | 7 | 8 | 12 | 5 | 7 | 5 | 6 | 1 | 0 | 1 | 0 | 0 | 1 | 84 |
| 0207 | 10 | 10 | 15 | 7 | 7 | 14 | 3 | 5 | 5 | 2 | 2 | 1 | 1 | 3 | 3 | 3 | 91 |
| 0206 | 10 | 16 | 11 | 6 | 14 | 7 | 6 | 4 | 6 | 4 | 0 | 3 | 3 | 1 | 0 | 5 | 96 |
| 0205 | 14 | 18 | 12 | 8 | 9 | 6 | 5 | 5 | 5 | 2 | 1 | 1 | 0 | 5 | 0 | 5 | 96 |
| 0204 | 11 | 18 | 8 | 14 | 8 | 4 | 8 | 3 | 1 | 1 | 1 | 0 | 2 | 2 | 0 | 3 | 84 |
| 0203 | 12 | 15 | 16 | 9 | 4 | 8 | 2 | 3 | 1 | 3 | 0 | 2 | 0 | 2 | 2 | 8 | 87 |
| 0202 | 17 | 20 | 14 | 11 | 2 | 5 | 5 | 3 | 2 | 0 | 0 | 0 | 0 | 3 | 2 | 9 | 93 |
| 0201 | 25 | 15 | 11 | 1 | 9 | 8 | 0 | 0 | 0 | 0 | 0 | 0 | 1 | 2 | 0 | 5 | 77 |
| 0131 | 23 | 17 | 10 | 7 | 3 | 4 | 2 | 0 | 0 | 0 | 0 | 0 | 0 | 2 | 3 | 13 | 84 |
| 0130 | 19 | 16 | 7 | 3 | 3 | 2 | 0 | 0 | 0 | 0 | 0 | 0 | 0 | 0 | 5 | 0 | 14 | 69 |
| 0129 | 14 | 7 | 3 | 5 | 3 | 2 | 3 | 0 | 0 | 0 | 0 | 1 | 0 | 2 | 15 | 19 | 74 |
| 0128 | 20 | 10 | 4 | 4 | 1 | 1 | 0 | 0 | 0 | 1 | 1 | 1 | 2 | 13 | 1 | 22 | 81 |
| 0127 | 14 | 11 | 5 | 4 | 4 | 0 | 2 | 0 | 1 | 0 | 1 | 5 | 3 | 12 | 0 | 14 | 76 |
| 0126 | 16 | 10 | 7 | 6 | 1 | 2 | 0 | 0 | 1 | 0 | 3 | 2 | 4 | 2 | 10 | 15 | 79 |
| 0125 | 18 | 8 | 6 | 3 | 3 | 1 | 1 | 1 | 0 | 1 | 3 | 1 | 6 | 9 | 8 | 18 | 87 |
| 0124 | 9 | 10 | 15 | 4 | 3 | 5 | 1 | 0 | 0 | 0 | 7 | 2 | 4 | 0 | 10 | 15 | 85 |
| 0123 | 10 | 13 | 6 | 1 | 1 | 1 | 1 | 0 | 1 | 1 | 3 | 2 | 5 | 15 | 0 | 16 | 76 |
| 0122 | 12 | 7 | 5 | 2 | 3 | 2 | 3 | 0 | 1 | 3 | 0 | 0 | 9 | 18 | 0 | 11 | 76 |
| 0121 | 9 | 11 | 4 | 1 | 2 | 2 | 1 | 1 | 0 | 3 | 1 | 1 | 11 | 15 | 1 | 8 | 71 |
| 0120 | 9 | 2 | 3 | 2 | 1 | 0 | 1 | 0 | 1 | 3 | 4 | 5 | 16 | 18 | 1 | 9 | 75 |
| 0119 | 6 | 0 | 1 | 0 | 1 | 2 | 0 | 0 | 1 | 3 | 7 | 13 | 15 | 14 | 1 | 16 | 80 |
| 0118 | 0 | 3 | 1 | 0 | 0 | 0 | 2 | 3 | 4 | 15 | 13 | 12 | 19 | 2 | 11 | | 85 |
| 0117 | 1 | 0 | 1 | 0 | 0 | 1 | 1 | 2 | 6 | 24 | 13 | 6 | 9 | 15 | 2 | 5 | 86 |
| 0116 | 0 | 0 | 0 | 0 | 1 | 2 | 2 | 6 | 23 | 18 | 10 | 19 | 7 | 3 | 1 | 1 | 93 |
| 0115 | 0 | 0 | 0 | 1 | 0 | 2 | 5 | 23 | 19 | 17 | 10 | 8 | 3 | 0 | 2 | 2 | 92 |
| 0114 | 0 | 0 | 1 | 1 | 0 | 4 | 17 | 43 | 16 | 5 | 3 | 4 | 2 | 0 | 0 | 0 | 96 |
| 0113 | 1 | 0 | 0 | 1 | 0 | 18 | 8 | 31 | 20 | 9 | 3 | 1 | 0 | 0 | 0 | 0 | 92 |

**Extended Data Fig. 5 | Importance of the spectrum. Analysis of the influence of light sources of different colors, sound, and eating start time on the egg-laying phase of hens.** For illumination with three RGB light sources, the following light intensities were chosen to ensure the same number of photons: R = 1.3 Lx, G = 6.0 Lx, and B = 0.22 Lx. The corresponding spectra are displayed in ED Fig. 7d. The experiment started on 2020-1-13 (0113). At first, a contrast experiment using green light was performed, which showed that green light at a light intensity of 6 Lx could change the egg-laying phase of hens about 1.5 h per day, which proved that the light intensity of 6 Lx was reasonably chosen. After this experiment, we were obliged to change our research plan due to the COVID-19 outbreak and kill all the hens as soon as possible. Therefore, Flocks A and B were no longer used for comparison. For the same number of photons, the effect of red light was faster than that of green light, while blue light hardly influenced the egg-laying phase. Furthermore, the total number of eggs laid under exposure to red light decreased unexpectedly. In the dark, music could not change the egg-laying rhythm. Under illumination with light of an intensity of 10 Lx, the egg-laying phase of hens did not shift about 6 h when the eating start time in the morning and evening was changed to noon and midnight (i.e., shift of the eating start time by 6 h). This shows that the eating start time has no influence on the egg-laying phase, which can be only changed by light. Furthermore, this suggests that light, rather than diet, is the main determinant of human reproduction when nutrition required for the metabolism is abundantly available, and light pollution is more harmful than food pollution (**M24**).

# Extended Data

| date ↑ | 0:00 | 1:30 | 3:00 | 4:30 | 6:00 | 7:30 | 9:00 | 10:30 | 12:00 | 13:30 | 15:00 | 16:30 | 18:00 | 19:30 | 21:00 | 22:30 | sum |
|---|---|---|---|---|---|---|---|---|---|---|---|---|---|---|---|---|---|
| 1106 | 15 | 17 | 9 | 4 | 5 | 1 | 1 | 3 | 0 | 0 | 2 | 1 | 4 | 3 | 11 | 10 | 86 |
| 1105 | 12 | 16 | 7 | 4 | 4 | 1 | 4 | 1 | 1 | 1 | 3 | 3 | 1 | 0 | 13 | 8 | 79 |
| 1104 | 9 | 11 | 7 | 4 | 2 | 0 | 4 | 3 | 2 | 1 | 1 | 3 | 5 | 2 | 13 | 10 | 77 |
| 1103 | 6 | 9 | 6 | 2 | 2 | 1 | 6 | 2 | 2 | 4 | 3 | 6 | 2 | 2 | 14 | 15 | 82 |
| 1102 | 10 | 7 | 7 | 1 | 6 | 0 | 5 | 2 | 5 | 4 | 5 | 4 | 1 | 6 | 13 | 12 | 88 |
| 1101 | 7 | 5 | 6 | 4 | 1 | 3 | 4 | 0 | 4 | 4 | 3 | 7 | 12 | 13 | 6 | 2 | 82 |
| 1031 | 9 | 9 | 6 | 1 | 1 | 4 | 2 | 2 | 3 | 4 | 11 | 8 | 8 | 9 | 9 | 8 | 94 |
| 1030 | 7 | 5 | 5 | 1 | 1 | 0 | 4 | 5 | 8 | 5 | 15 | 8 | 5 | 1 | 12 | 2 | 84 |
| 1029 | 11 | 5 | 3 | 3 | 0 | 2 | 2 | 4 | 6 | 11 | 9 | 7 | 7 | 7 | 2 | 5 | 84 |
| 1028 | 6 | 3 | 2 | 2 | 0 | 0 | 2 | 8 | 7 | 12 | 12 | 8 | 0 | 0 | 12 | 3 | 77 |
| 1027 | 2 | 1 | 1 | 0 | 2 | 0 | 4 | 7 | 5 | 14 | 17 | 10 | 6 | 6 | 3 | 5 | 83 |
| 1026 | 0 | 4 | 0 | 2 | 1 | 2 | 1 | 10 | 14 | 25 | 6 | 7 | 9 | 0 | 3 | 4 | 88 |
| 1025 | 0 | 2 | 2 | 1 | 4 | 0 | 7 | 10 | 20 | 14 | 4 | 9 | 7 | 4 | 0 | 2 | 86 |
| 1024 | 1 | 1 | 0 | 1 | 4 | 3 | 8 | 13 | 16 | 8 | 8 | 11 | 4 | 3 | 5 | 0 | 86 |
| 1023 | 0 | 0 | 1 | 1 | 2 | 3 | 1 | 5 | 18 | 15 | 10 | 15 | 2 | 5 | 2 | 2 | 83 |
| 1022 | 0 | 1 | 0 | 2 | 0 | 1 | 8 | 14 | 12 | 14 | 9 | 4 | 4 | 8 | 0 | 1 | 78 |
| 1021 | 0 | 0 | 1 | 1 | 3 | 2 | 6 | 15 | 12 | 15 | 9 | 7 | 8 | 5 | 0 | 1 | 85 |
| 1020 | 0 | 1 | 2 | 1 | 1 | 0 | 6 | 14 | 19 | 11 | 12 | 8 | 6 | 0 | 1 | 1 | 83 |

a

| date ↑ | 0:00 | 1:30 | 3:00 | 4:30 | 6:00 | 7:30 | 9:00 | 10:30 | 12:00 | 13:30 | 15:00 | 16:30 | 18:00 | 19:30 | 21:00 | 22:30 | sum |
|---|---|---|---|---|---|---|---|---|---|---|---|---|---|---|---|---|---|
| 1106 | 0 | 1 | 0 | 1 | 2 | 7 | 10 | 10 | 11 | 13 | 7 | 3 | 8 | 0 | 4 | 0 | 77 |
| 1105 | 1 | 1 | 1 | 0 | 4 | 6 | 13 | 12 | 11 | 17 | 7 | 4 | 3 | 2 | 1 | 0 | 83 |
| 1104 | 0 | 3 | 4 | 0 | 1 | 6 | 14 | 7 | 11 | 13 | 8 | 4 | 3 | 3 | 1 | 0 | 78 |
| 1103 | 0 | 2 | 2 | 1 | 0 | 1 | 11 | 6 | 13 | 16 | 9 | 6 | 2 | 0 | 1 | 0 | 70 |
| 1102 | 2 | 1 | 3 | 1 | 4 | 11 | 10 | 14 | 13 | 9 | 4 | 3 | 2 | 2 | 2 | 2 | 83 |
| 1101 | 2 | 3 | 3 | 6 | 3 | 6 | 10 | 10 | 8 | 13 | 4 | 2 | 8 | 3 | 1 | 0 | 82 |
| 1031 | 4 | 4 | 3 | 5 | 4 | 5 | 11 | 13 | 13 | 6 | 6 | 3 | 1 | 2 | 4 | 1 | 87 |
| 1030 | 2 | 3 | 4 | 2 | 7 | 12 | 12 | 11 | 13 | 4 | 4 | 4 | 3 | 0 | 2 | 3 | 86 |
| 1029 | 2 | 4 | 4 | 11 | 6 | 12 | 6 | 8 | 6 | 8 | 2 | 4 | 6 | 2 | 2 | 2 | 85 |
| 1028 | 5 | 8 | 9 | 9 | 12 | 6 | 0 | 7 | 4 | 4 | 3 | 7 | 5 | 3 | 0 | 2 | 84 |
| 1027 | 9 | 6 | 9 | 9 | 9 | 11 | 7 | 1 | 6 | 5 | 6 | 1 | 2 | 2 | 1 | 3 | 87 |
| 1026 | 6 | 9 | 11 | 15 | 10 | 6 | 3 | 5 | 3 | 4 | 1 | 2 | 1 | 1 | 4 | 8 | 89 |
| 1025 | 5 | 15 | 13 | 11 | 6 | 6 | 3 | 3 | 1 | 2 | 3 | 1 | 0 | 0 | 5 | 12 | 86 |
| 1024 | 10 | 24 | 10 | 10 | 4 | 3 | 3 | 2 | 1 | 2 | 0 | 1 | 2 | 3 | 2 | 12 | 89 |
| 1023 | 13 | 19 | 9 | 9 | 1 | 3 | 2 | 5 | 1 | 2 | 0 | 0 | 1 | 0 | 10 | 11 | 86 |
| 1022 | 20 | 14 | 5 | 8 | 8 | 2 | 1 | 2 | 0 | 1 | 0 | 0 | 3 | 3 | 8 | 17 | 92 |
| 1021 | 20 | 14 | 8 | 5 | 3 | 0 | 4 | 1 | 2 | 0 | 1 | 0 | 1 | 4 | 8 | 11 | 82 |
| 1020 | 20 | 6 | 6 | 8 | 4 | 6 | 3 | 1 | 0 | 0 | 0 | 3 | 1 | 6 | 4 | 18 | 86 |

b

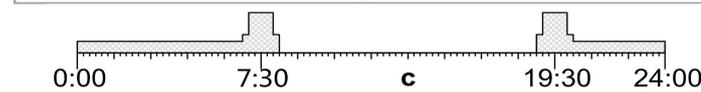

0:00        7:30        **c**        19:30        24:00

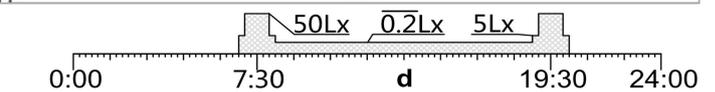

0:00        7:30        **d**        19:30        24:00

**Extended Data Fig. 6 | Brightness had no special effect on the egg-laying rhythm of hens**. **a, b |** The experiment started on 2019-10-20 (1020), and a single LED bead was used as the light source (spectrum in ED Fig. 8a). **c, d |** Illumination phases of Flocks A and B. With increasing distance between the hens and the LED bead, the illumination at the hens' eyes decreased from 0.3 to 0.1 Lx (ED Fig. 7b) with an average value of 0.2 Lx. The brightness of the LED bead was estimated to be above $10^6$ cd/m$^2$. For comparison, the brightness of a TV screen is about 300 cd/m$^2$, while that of the full moon is about 6000 cd/m$^2$. Compared with Fig. 1 in the main text, the change rate of the egg-laying phase was basically the same. However, the perceived brightness in Fig. 1 was estimated to be below 10 cd/m$^2$, which is much less than $10^6$ cd/m$^2$. Therefore, in contrast to illumination, brightness is not a decisive factor affecting the egg-laying phase of hens. If this result can be transferred to humans, the brightness of the moon surface may not be a key factor affecting human rhythms (e.g., menstruation; also see **M20**).

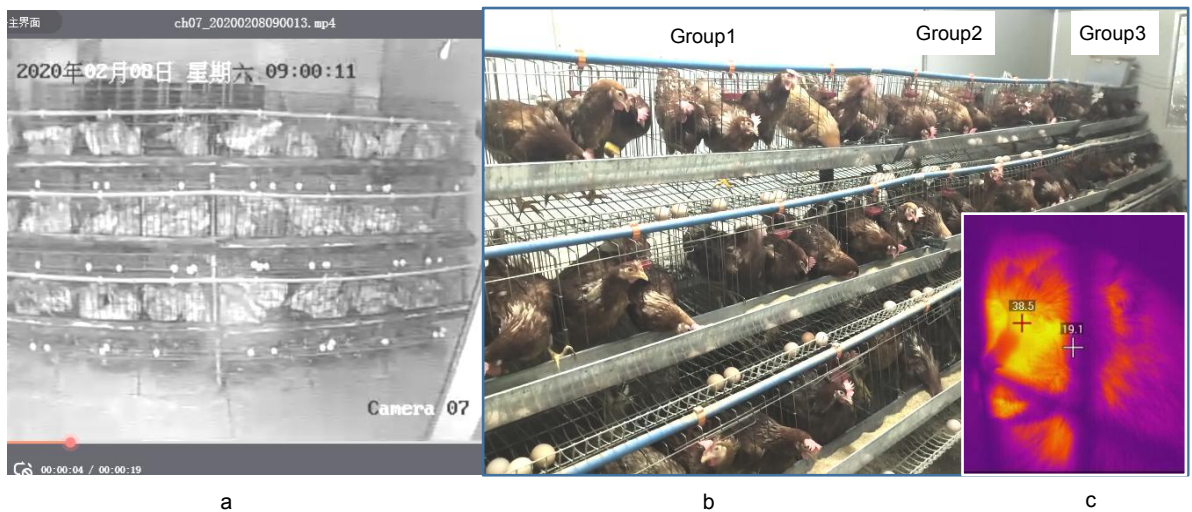

a        b        c

**Extended Data Fig. 7 | "Egg-counting" method for the study of the circadian rhythm of hens.** The thermal imaging camera C640Pro (Wuhan Guide Infrared Co., Ltd.) was used to record the variation in body temperature during ovulation in monkeys and hens. **a |** In the dark, it was impossible to change the egg-laying phase with loud voice (screenshot from ED Fig. 6a on 2020-02-08), and the hens still slept with loud music. **b |** Photograph of the arrangement of Flock B, which included three groups of hens. **c |** In an investigation of the relationship between menstruation and light on a Chinese macaque island in Jiangxi, it was impossible to find menstrual blood in monkeys. Therefore, we were obliged to change our research plan, and we purchased a high-precision infrared thermal camera and tried to determine the ovulation date by monitoring the difference in the local body temperature. However, we discontinued this method, as it was too expensive. We also used this camera to study the temperature variation during the circadian rhythm of hens, which showed that this method was more difficult than our egg-counting method.

# Extended Data

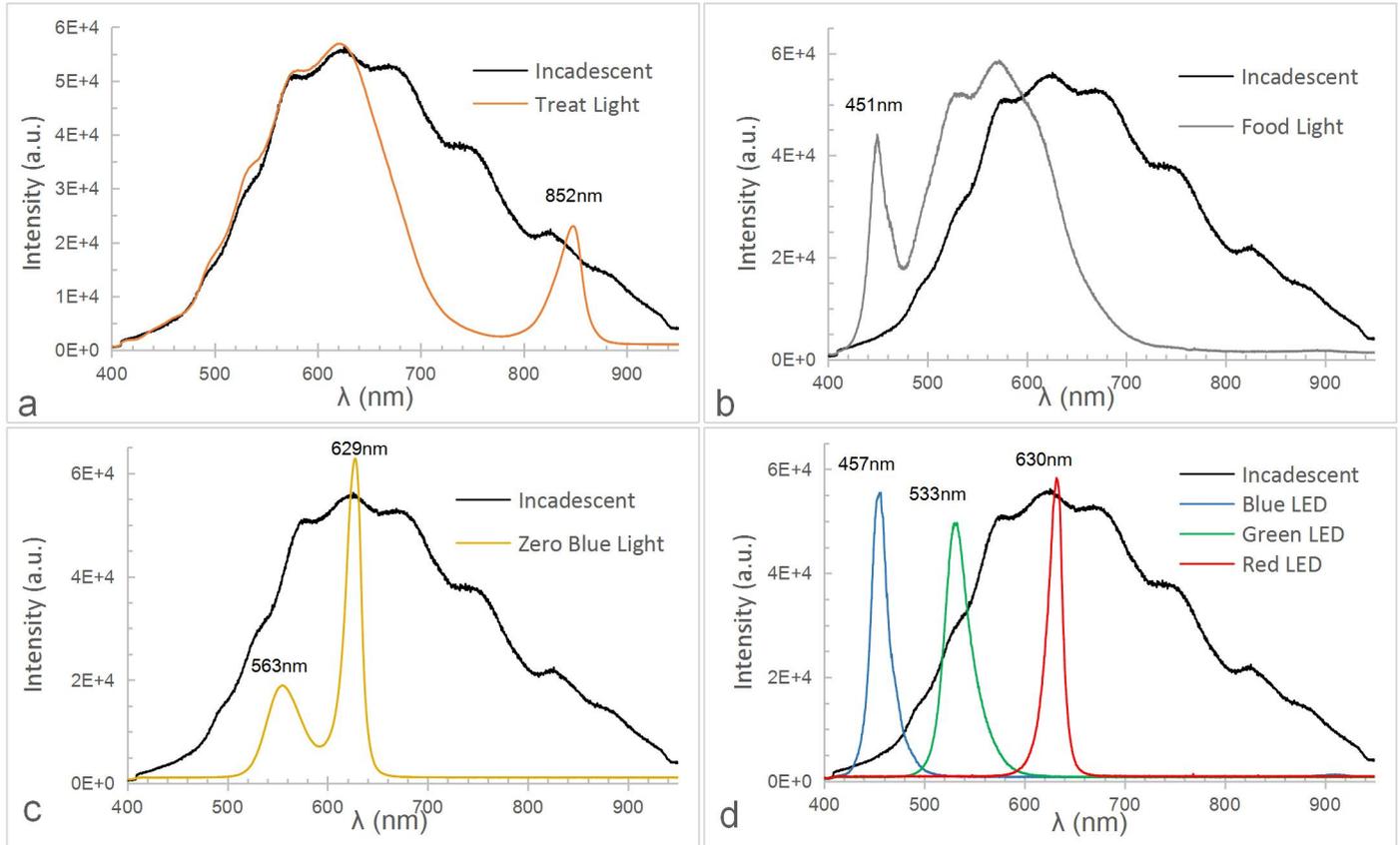

**Extended Data Fig. 8 | The used light source was precise and controllable by DC. a |** The spectrum of a self-made LED light source that accurately imitates the incandescent spectrum at 2700 K (measured with an Ocean Optics spectrometer USB2000+ and referenced to incandescent light) with additional 852 nm infrared light. This LED light source is characterized by DC dimming, purple LED (405 nm) excitation, full visible spectrum, and supplementary infrared light. The use of this self-made LED lamp eliminates all possible inaccuracies from the light source that occurred in similar kinds of reported experiments. **b |** Spectrum of a self-made high-quality LED light source during feeding with a color temperature of about 4500 K and color-rendering index Ra > 90 but without strobe. **c |** Spectrum of a self-made zero-blue LED light source (golden light). **d |** Spectrum of a red LED (630 nm), green LED (525 nm), and blue LED (460 nm) used in this experiment.

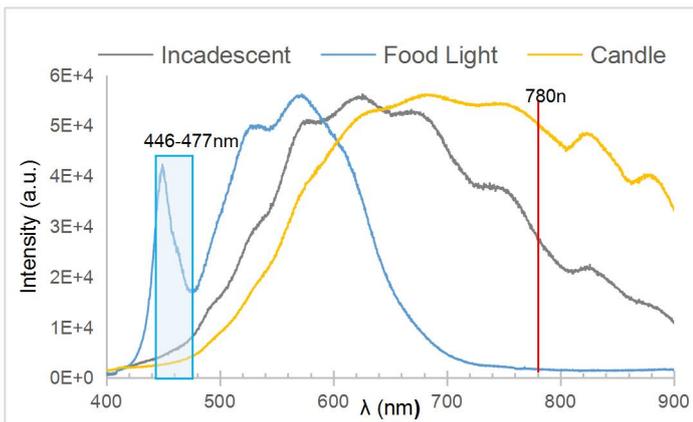

**Extended Data Figure 9 |** The blue light ratio of different light sources

**Extended Data Table 1 | Candlelight was used as a control to intuitively evaluate the proportion of the blue light content (blue ratio) in various light sources.** The whiter light is, the more blue light it contains. Ref. 3 showed that blue light with a wavelength of 446–477 nm most likely inhibits the pineal secretion of melatonin in humans. Therefore, the ratio between the spectral area of the visible band at 446–477 nm and that at 400–780 nm represents the proportion of blue light of the light source. Intuitively, the blue light content of other light sources can be calculated by taking candlelight as a reference and setting its blue ratio to 1. For example, the blue ratio of the LED lamp varies from 2.6 to 15. The early poor LED lamp has a high blue ratio of 27, while the smartphone, while displaying white, can have a blue ratio of as high as 19.

|  | Candle | Incandescent light bulb | Treatment lamp | Feeding lamp |
| --- | --- | --- | --- | --- |
| Blue light ratio | 1 | 1.8 | 2.6 | 12 |

1Institute of material Science, Nanchang University, 235# NanJinDong Road, Nanchang, JIangxi 330046,P.R.China. 2 School of Chemistry and Materials, Shangluo College, Shangluo, Shaanxi，P.R.China,
†Present addresses:  fwq@ncu.edu.cn , cpyang2022@sinano.ac.cn

# The relationship between light and the determinants of fertility

In this Extended Data file, we tried to correlate all determinants of fertility listed in a recent review[28] with light according to the hypothesis put forward in the main text, which is in turn confirmed by a positive correlation. The general idea is based on: 1) Medicine: the non-visual effects of light can affect human's hormone secretion, which affects their values and behavior. In humans, cortisol secretion and sex hormone secretion are associated with exposure to light with blue light content. Blue light at night could act as a stimulant or hormone drug. 2) Evolution and philosophy: human consciousness may originate from the pursuit of light in darkness. Light may be the source of excitement, and darkness may be the source of desire. If humans were exposed to light day and night, they would become desireless. 3) Philosophy: the advent of the 5G era may have drawbacks, which need to be considered. 4) Optics: the non-visual effect of light experienced by people nowadays at night is 120,000 times stronger than that people 70 years ago were exposed to. The general state of modern people, including their sexual state, is significantly different from that of people who lived 70 years ago. Modern people may be in a state of madness, which should be considered when their decisions are evaluated, and it is worth studying whether modern people have the ability to correct their deviations. 5) Nowadays, all young people are sexually precocious. 6) Evidence from strategies applied and effects observed at a pig farm.

Below on the left, the fertility factors reported in the review are listed[28], and the association of these factors with the exposure to blue light is presented on the right. However, we admit that some of the links may be vague.

## Micro-Level Determinants of Fertility (at the individual and/or couple level)

| | |
|---|---|
| Individual's values and preferences | Values and preferences of individuals are affected by their hormone levels and thus by light. |
| Fertility intentions | The "ideal family size" may be related to "maternal love" or endocrine secretion, which is related to light at night. The three intention factors of "perceived costs and benefits", "influence of close friends and relatives", and "perceived control over behavior" could also be related to light. The reproductive behavior may not only be the result of a reasonable rational deliberation but also that of an automatic unconscious processing. This unconscious processing may be related to maternal love, which is affected by the exposure to light. |
| Partner and Partnership | An "increased frequency of having several partners before the first child" indicates that precocious modern young men and women are in a state of sexual excitement due to exposure to the blue component in light at night. The partner's fertility intentions also play an important role in the realization of an individual's intentions. More and more young people 'retreat from marriage', because light is like a drug that has altered their mental state and hormonal state. |
| Births out of wedlock | Cohabitation is affected by hormone levels, so it is affected by light. Moreover, cohabitation is associated with a lower probability of childbearing. |
| High separation and divorce level | Young men and women are in a state of sexual excitement due to the exposure to light with blue light content at night, resulting in a partnership with low quality or instability. Without "maternal love" or "sense of responsibility", they focus on the fact that children may raise the costs of a separation. |
| Work–family conflict of woman | Gendered division of labor: compared with previous generations, modern people's life is very convenient, as the amount of housework labor has been greatly reduced. In China, in the 1960s and 1970s, families had many children, and the amount of housework was very large. However, women received equal pay for the same work as men, and most of them had grandmothers to look after the children. Therefore, the conflict between work and family is likely to be an excuse, and its essence is the lack of maternal love and sacrifice in the entire society, including the grandmothers. Maybe grandmothers also despise their sexually open daughters-in-law, which maybe also ascribed to the influence of light. |
| Stepfamily and Fertility | The increase in unstable families and multiple partnerships also resulted in an increase in the research on stepfamily fertility. It is difficult for these families to get the best support and blessing from the grandparents, and it is difficult for couples to be completely united. The instability of partnerships may result from exposure to too much light. |
| Woman's education and career planning | Fertility is more costly for higher-income mothers. While these mothers received a longer education, they were exposed to more blue light at night, so their "maternal love" may be less. Parents with a higher income value their children's education quality, and according to the quality-quantity tradeoff, the quantity will be reduced. Furthermore, longer education times result in more exposure of these children to blue light. |
| Economic and Employment Uncertainty | Economic uncertainty is naturally linked to the postponement of parenthood. But in welfare states, young adults can receive protection to shelter from uncertainty to some extent. This raises the questions why young adults think that the state's social security is not sufficient to have more children if food and clothes are supplied, why their parents do not do their best to support them, and why young people lose their jobs. These questions may be answered by the exposure to blue light, which alters the hormone production of young people and the whole society. As a result, people lose their animal instinct to have more offspring to form the next generation. People do not have enough motherly love and dedication, so the whole society may become greedy and selfish. Some people may maximize their own interests and increase the production while minimizing the number of employees. |

# Supplementary Information

| | |
|---|---|
| Intergenerational Transmission of Values and Behavior | It is assumed that parents transmit family values, preferences, and attitudes, as well as degree of religiosity. However, the inheritance of childbearing is much weaker today, as older generations seem to have lost confidence in the new technology of child rearing, and they seem to lack enough maternal love to worship the idea of having a big family. Older generations often look down on young wives who were sexually open, and young people in a state of excitement also have no patience to listen to the previous generation. Regarding the reproductive religion and considering that blue light may act as a drug, we cannot expect a drug addict to have normal religious beliefs. If the whole society were in a state of sexual excitement, there would be no normal religious belief in reproduction. |
| Biological and genetic factors | An interdisciplinary area of fertility research includes behavioral and molecular genetics, neuroendocrinology, and evolutionary theory. To the best of our knowledge, no attention has been paid yet to the effects of artificial light in the last hundred years. |
| Childbearing, well-being, and social esteem | It is a human instinct to raise more children, which brings happiness and honor as well as the hope of being respected by society. However, too much blue light has increased the threshold of happiness, and young people addicted to mobile games may be far happier playing mobile games than being parents. Too much blue light causes a lack of basic maternal love, and people will only care about their own play life but no longer about the fertility of their neighbors. As a result, having more children does not result in more social respect, and a home with more children will even be considered as a lower-class household. |

## Meso-Level Determinants of Fertility

Meso-level means individuals that are social actors who make decisions and act while being embedded in a web of social relationships with kin and peers.

| | |
|---|---|
| Social Interaction | Individuals look to coworkers or siblings, for example, as a key source of social learning to see how and whether they successfully navigate the combination of having children with a career. Due to the ubiquity of night lighting in advanced countries, even the few who want to have more children are not immune to the influence of the coworkers that are in a state of excitement. |
| Place of Residence | Fertility differentials between small towns and rural areas or between urban areas and suburbs is an important evidence of the effect of light on human reproduction. The fertility rates of second-generation immigrants quickly became similar to those of the local population, which is also an important evidence for the influence of light because they share the same nighttime light environment. |
| Social Capital | Social capital is defined as the resources that individuals have access to via personal relationships. It includes goods, information, money, capacity to work, influence, power, or active help. Next to economic and cultural resources, social capital can also shape fertility decision-making. Multipartnered fertility, through the difficulties of maintaining kin networks, lowers the financial, housing, and child-care support to mothers. As a result, too much blue light at night can cause young people to act strangely and not get the full support of friends and family. |

## Macro-Level Determinants of Fertility

| | |
|---|---|
| Economic Trends | It has been reported that a higher GDP is associated with higher fertility, but other reports suggested that economic upturn is related with an increased employment of women, which makes having children more expensive during times of economic prosperity. Therefore, fertility trends are likely to be countercyclical. In Taiwan and South Korea, the fertility declines with an increasing GDP. We believe that a higher GDP results in more night work and better lighting conditions. More blue light at night will lead to a decrease in fertility, which is observed in basically all countries in the world. |
| (Un)employment Trends | The higher the unemployment, the lower the fertility. Too much light will bring people into a state of excitement and madness, resulting in a lack of vision and dedication, and people will voluntarily and frantically work day and night and be exposed to more light in the meantime. In this way, one employee can do the work of others, one factory can crush several factories, and the output of the whole society in one year can meet the needs of several years. Therefore, blue-light hazard may affect employment. Furthermore, people without a job rely on watching TV or playing online games to kill time, but they also receive more light to enter a state of drug use and decadence. These people lose hope for future and will not consider having children. |
| Policy Measures Fiscal Incentives | Policy measures have an impact on the timing of parenthood, as well as on the number of children. Public availability of childcare, generous child allowances, parental-leave allowance, and other social security systems have been discussed as determinants of fertility. In advanced countries, these policies did not recover the fertility, as humans in a state of excitement are extremely greedy. Policy makers are also overexposed to blue light and lack of courage and sacrifice necessary to correct the wrong policy. For example, the result of a generous pension may be that grandparents have no interest in taking care of their children. However, policy makers usually just want to meet the demand of voters in the excited state of greed. Regardless of these requirements, these wrong policies may be harmful to the healthy development of human beings in the future. |
| Welfare Regimes | The welfare regimes of developed countries are quite good compared with those of developing countries, which did unexpectedly not lead to a significant increase in fertility. We think that this is partly because humans tend to ignore what they already have. As people's perception of light is logarithmic in optics, they need a tenfold stronger stimulus to get excited again. However, the better the welfare regimes are, the less people want. On the other hand, with good welfare regimes, people have more time for nightlife, and too much light makes active young people sexually excited, which is not conducive to the establishment of stable families. Too much light also makes inactive young people addicted to games, which is tantamount to being in a drug state. |

# Supplementary Information

| | |
|---|---|
| Value and Attitude Changes | "Ideational changes, which mainly consist of the rejection of institutional control, accentuation of individual autonomy, and the rise of self-realization needs, are the driving forces of new family arrangements and behaviors, among which fertility postponement, reduced number of children, or childlessness have developed since the 1960s". These changes coincided with the spread of fluorescent lights and color TV, which increased indoor illumination by nearly 10 times at night. People in developed countries first entered a state of sexual excitement. Their pursuit of individual liberation may be not healthy and normal, and this change in value and attitude is fatal for their fertility. Society's tolerance of homosexuality, celibacy, and voluntary childlessness may be a manifestation of the lack of a sense of justice caused by the overexposure to light, but civilization requires justice. |
| Historical and Cultural Continuities | Today, in South Korea, Taiwan, and China, the tradition of having more children has disappeared. The fertility of migrants is often similar to that of their place of destination. Therefore, the influence of historical and cultural traditions on fertility is difficult to continue. Major religious denominations generally advocate fertility, but similar to drug addicts who do not follow cultural and historical traditions, modern humans, who are exposed to blue light at night for long periods of time, cannot restrain their reproductive behavior by tradition. |
| Contraceptive and Reproductive Technologies | Reproductive technology is expensive, and its contribution to increasing the fertility very limited. It may even mislead people, who may postpone having children to a later point of time when their fertility may have already decreased. Caesarean sections, which constitute up to 50% of births in some parts of China, also severely limit the fertility. The impact of various contraceptive technologies and abortions on fertility is even more devastating. When we realize that young people's pursuit of individual freedom may be in essence a pursuit of sexual excitement and not a noble act, and that policy makers themselves may be in a state of excitement and selfishness, fearful of being accused of restricting civil liberties, we could significantly improve fertility by largely limiting the use of contraceptive technologies and blue light at night. |

## Discussion

In this article, we have correlated blue light with all fertility factors listed in the review literature[28]. In some reports, this association is only presented as a hypothesis. For example, how short-term excitement and greed, as well as the final low desire and decadence, can be explained by the exposure to excessive light. In this article, we developed the hypothesis that blue light acts as a drug or hormone drug to solve this problem. Due to the logarithmic response of humans to light intensity, the threshold for new desires increased 10-fold. If the opposite is true, people enter a decadent state, namely a state of low desire. Therefore, human greed is insatiable, welfare policies that encourage childbearing have had a limited effect. Another important hypothesis formulated in this article is that people's values and dedication are related to their hormone levels, and thus related to blue light. Woman's love and man's righteousness and courage are correlated with the balance between brightness and darkness. It is an animal instinct to have more offspring, and disruption of this balance weakens the animal instinct.

An important point of this article is to analyze the nature of the pursuit of individual freedom, and that modern humans are in a state of sexual arousal due to excessive nighttime light, which is not considered noble. Policy makers may also be in an abnormal state, and the policies designed to increase the fertility of the people may not work well in the end.

Civilization implies that justice and dedication are practiced, and an appropriate human fertility may be the most important indicator of a healthy civilization. In the extremal case of zero fertility, no new generation is formed, which initiates the end of civilization. With the advent of the 5G era, the decline of fertility is expected to become even more severe, and prevention is required rather than focusing on economic interests.